\documentclass[prd,twocolumn,showpacs,reprint,preprintnumbers,nofootinbib,superscriptaddress]{revtex4-1}

\input{header}
\usepackage{placeins}
\begin{document}

\preprint{}

\title{Detecting LHC Neutrinos at Surface Level}

\author{Akitaka Ariga \orcid{0000-0002-6832-2466}}
\affiliation{Albert Einstein Center for Fundamental Physics,
Laboratory for High Energy Physics, University of Bern,
Sidlerstrasse 5, CH-3012 Bern, Switzerland}
\affiliation{Department of Physics, Chiba University,
1-33 Yayoi-cho Inage-ku, 263-8522 Chiba, Japan}

\author{Steven Barwick \orcid{0000-0003-2050-6714}}
\affiliation{Department of Physics and Astronomy, University of California, Irvine, CA 92697 USA}

\author{Jamie Boyd \orcid{0000-0001-7360-0726}}
\affiliation{CERN, CH-1211 Geneva 23, Switzerland}

\author{Max Fieg \orcid{0000-0002-7027-6921}}
\affiliation{Department of Physics and Astronomy, University of California, Irvine, CA 92697 USA}

\author{Felix Kling \orcid{0000-0002-3100-6144}}
\affiliation{Deutsches Elektronen-Synchrotron DESY, Notkestr.~85, 22607 Hamburg, Germany}

\author{Toni M\"akel\"a \orcid{0000-0002-1723-4028}}
\email{tmakela@uci.edu}
\affiliation{Department of Physics and Astronomy, University of California, Irvine, CA 92697 USA}

\author{Camille Vendeuvre}
\affiliation{CERN, CH-1211 Geneva 23, Switzerland}

\author{Benjamin Weyer}
\affiliation{CERN, CH-1211 Geneva 23, Switzerland}

\begin{abstract}

The first direct detection of neutrinos at the LHC not only marks the beginning of a novel collider neutrino program at CERN but also motivates considering additional neutrino detectors to fully exploit the associated physics potential. As the existing forward neutrino detectors are located underground, it is interesting to investigate the feasibility and physics potential of neutrino experiments located at the surface-level. A topographic desk study is performed to identify all points at which the LHC's neutrino beams exit the earth. The closest location lies about 9~km east of the CMS interaction point, at the bottom of Lake Geneva. Several detectors to be placed at this location are considered, including a water Cherenkov detector and an emulsion detector. The detector designs are outlined at a conceptual level, and projections for their contribution to the LHC forward neutrino program and searches for dark sector particles are presented. However, the dilution of the neutrino flux over distance reduces the neutrino yield significantly, necessitating large and coarse detector designs. We identify the experimental challenges to be overcome by future research, and conclude that at present the physics potential of surface-level detectors is limited in comparison to ones closer to the interaction point, including the proposed Forward Physics Facility.
\end{abstract}

\maketitle 
\clearpage

\section{Introduction} 
\label{sec:introduction}

The observation of neutrinos produced at the Large Hadron Collider (LHC) by the FASER~\cite{FASER:2023zcr} and SND@LHC~\cite{SNDLHC:2023pun} experiments has initiated a novel forward neutrino program at the European Center of Nuclear Research (CERN). Collider neutrino measurements will probe proton and nuclear structure and serve to constrain parton distribution functions (PDFs)~\cite{Cruz-Martinez:2023sdv}, test gluon recombination~\cite{Bhattacharya:2023zei}, intrinsic charm~\cite{Maciula:2022lzk}, and provide crucial input to resolve outstanding questions in astroparticle physics~\cite{Anchordoqui:2022fpn, Sciutto:2023zuz, Bai:2022xad}. The experiments are also sensitive to a variety of effects beyond the Standard Model (SM)~\cite{Feng:2017uoz}. 
        
To exploit this physics potential, continuing and expanding this physics program in the high-luminosity LHC era is essential. Both the FASER~\cite{Boyd:2882503} and SND@LHC~\cite{Abbaneo:2895224, Abbaneo:2909524} collaborations plan to upgrade their detectors in their current locations for operation in LHC Run~4. However, the existing LHC infrastructure in which they are located only offers limited tunnel space, restricting the possible size and target mass for neutrino detectors. For this reason, the Forward Physics Facility (FPF), which would house several experiments with larger dimensions in a purpose-build cavern, has been proposed~\cite{Anchordoqui:2021ghd, Feng:2022inv, Adhikary:2024nlv}. 
All of these proposed detectors are to be placed underground relatively close to the LHC interaction point (IP) producing the neutrino beams. However, considering the limited space in the existing tunnels and the construction cost of a new cavern, it is intriguing to consider the alternative possibility of placing novel experiments at the surface-level emergence points of the neutrino beams and to assess whether such experiments can probe the same physics as their closer-to-IP counterparts.

We have conducted a topographic and bathymetric desk study to identify the locations at which the projected lines of sight (LOS) from the IPs reach the surface of the Earth, whose results are presented in \cref{sec:exitpoints}. These exit points are found to be between 9 and 183~km from the IPs. Since the beam intensity scales inversely with the squared distance to the IP, a detector placed 10~km away from the IP is exposed to a 400 times smaller flux than the operating FASER and SND@LHC experiments. Hence, matching or exceeding the event rates of the underground near detectors necessitates surface detectors with kiloton target masses. The highest-energy LHC neutrinos are strongly collimated and produced parallel to the beamline~\cite{DERUJULA199380,Park:2011gh,FASER:2019dxq,Kling:2021gos}, with a transverse spread of a few meters at a site 10~km away from the IP, implying that detectors should be elongated along the beam's path. To build the needed large volume detectors, more cost-effective detector technologies need to be considered. Constructing the physics case of such a detector thereby requires precise knowledge of the beam location, placing it as close to a high-luminosity IP as possible, and assessing the effects of the possibly limited experimental resolution of the chosen technologies on the projected constraining power of the experiment.

\begin{figure*}[t]
\centering
\hspace{4mm}\includegraphics[width=0.89\textwidth,trim={0mm 1mm 1mm 0mm},clip]{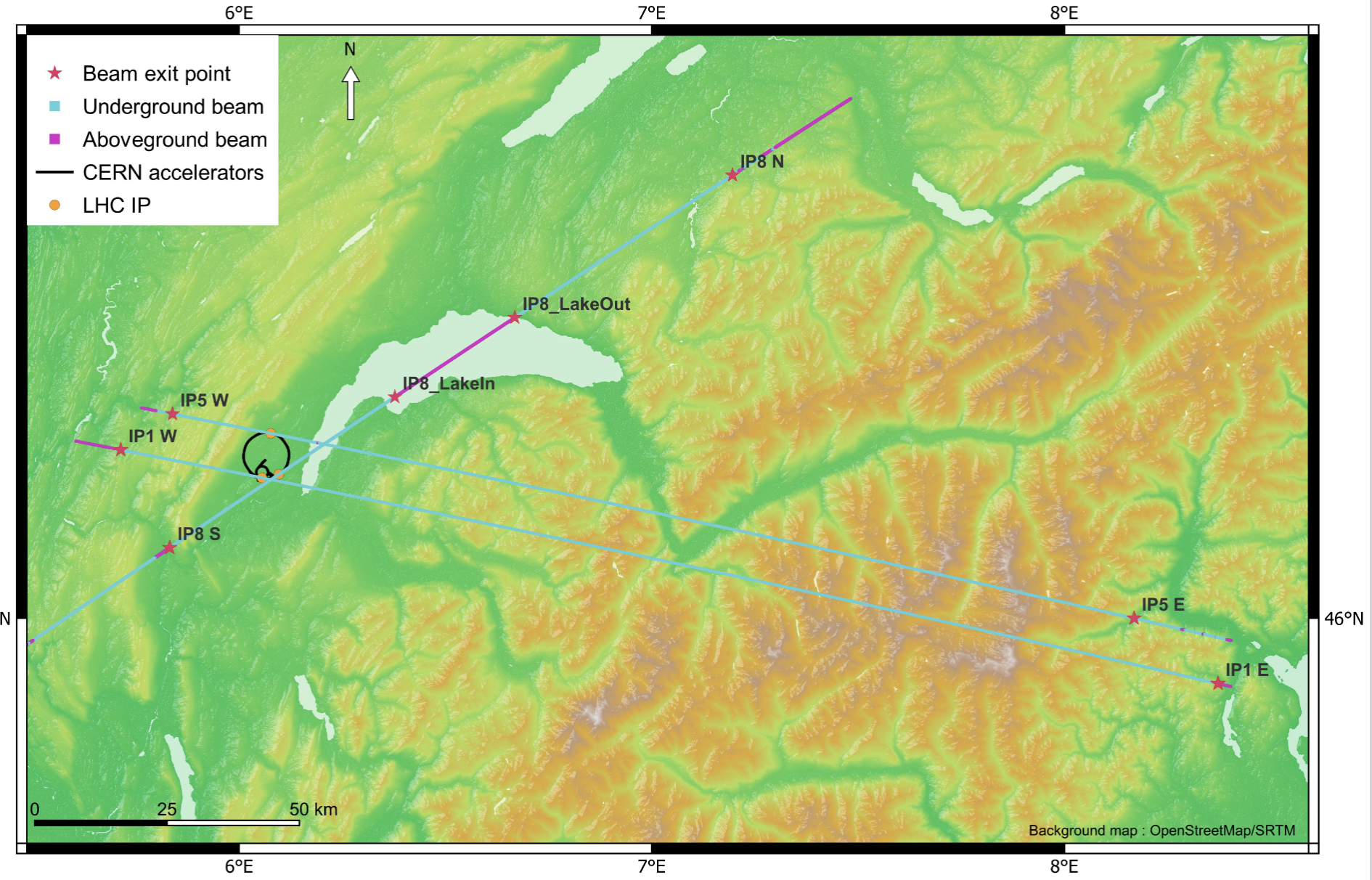}
\caption{The projected locations where the LOS from each IP reaches the surface of the Earth. The LOS is estimated to be above the lakebed between IP8\_LakeIn and IP8\_LakeOut. This location is referred to as IP8L in \cref{table:exitPointTable}.}
\label{fig:exitPointMap}
\end{figure*}

\begin{figure*}[t]
\centering
\includegraphics[width=0.94\textwidth,trim={0mm 8mm 1.5mm 8mm},clip]{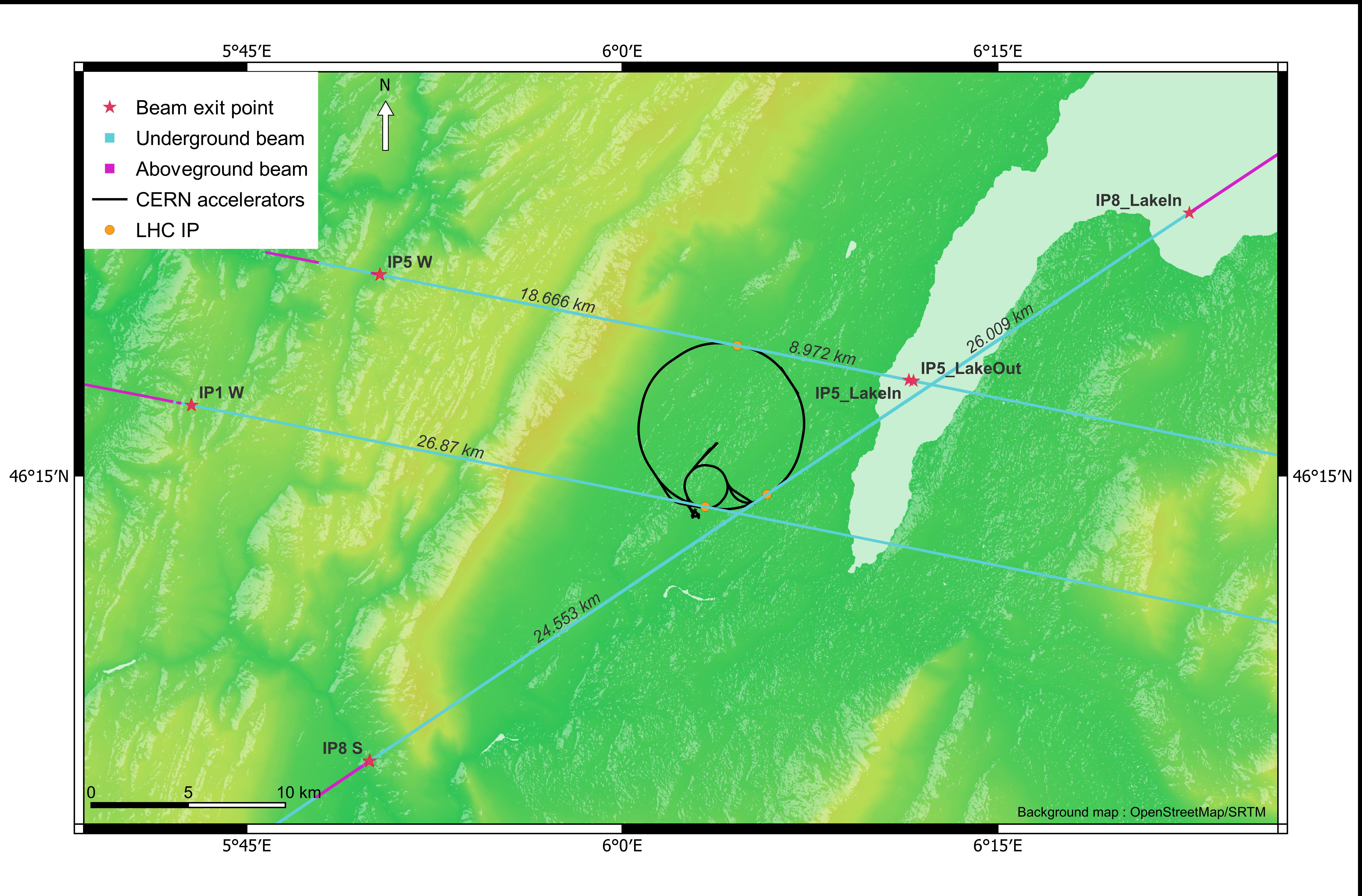}
\caption{A zoomed in version of~\cref{fig:exitPointMap}, showing the LOS exit points within 30km of the respective IPs. 
The LOS is estimated to be above the lakebed between IP5\_LakeIn and IP5\_LakeOut. This location is referred to as IP5L in the text.}
\label{fig:exitPointMapZoom}
\end{figure*}

\newpage
~\\

We discuss the optimal placement and alignment, potential performance, expected event rates and physics case for such detectors at a conceptual level. Particularly, the emergence point closest to one of the high-luminosity IPs, located about 9~km east of IP5 hosting the CMS experiment~\cite{Chatrchyan:2008zzk}, is identified as the most promising option. There, the neutrino beam passes through the bottom of Lake Geneva, which could be instrumented with, for example, a submerged emulsion detector or a water Cherenkov detector, with some potential to augment the LHC forward neutrino and dark sector physics program.

\begin{table*}[t!]
\setlength{\tabcolsep}{7pt}
\caption{Details of each LOS exit point considered, showing the expected luminosity to be delivered to the IP in the HL-LHC and the distance of the exit point from the corresponding IP. The flux relative to that for IP5L is estimated taking into account the distance from the IP and the expected luminosity. Also shown for comparison are the location of the operating LHC neutrino experiments as well as the FPF.}
\label{table:exitPointTable}
\begin{tabular}{c c c c c}
\hline
\hline
  IP/Side
    & luminosity 
    & distance 
    & relative flux
    & comment
    \\
\hline
  IP1W
    & 3000 fb$^{-1}$
    & 26.9~km
    & 0.1
    & in Jura mountains
    \\
  IP1E
    & 3000 fb$^{-1}$
    & 183~km
    & 0.0025 
    & very far
    \\
  IP5W
    & 3000 fb$^{-1}$
    & 18.7~km
    & 0.25
    & in Jura mountains
    \\
  IP5L
    & 3000 fb$^{-1}$
    & 9~km
    & 1
    & in Lake Geneva
    \\
  IP5E
    & 3000 fb$^{-1}$
    & 166~km
    & 0.0029
    & very far
    \\
  IP8L
    & 300--600 fb$^{-1}$
    & 26~km
    & 0.0125--0.025 
    & in Lake Geneva
    \\
  IP8S
    & 300--600 fb$^{-1}$
    & 24.6~km
    & 0.0133--0.0266
    & in Jura mountains
    \\
\hline
  FASER/SND 
    & 3000 fb$^{-1}$
    & 480~m
    & 351
    & TI12/TI18
    \\
  FPF
    & 3000 fb$^{-1}$
    & 620~m
    & 210
    & purpose-built cavern
    \\
\hline
\hline
\end{tabular}
\end{table*}

The paper is organized as follows. \cref{sec:exitpoints} discusses the results of the desk study and identifies the locations suitable for surface-level detector placement.  \cref{sec:detectors} introduces the proposed experimental setups and their locations, for which the event rates and predicted neutrino spectra are given in \cref{sec:spectra}. The physics case is presented in \cref{sec:applications}, and conclusions in \cref{sec:conclusions}.

\section{Surface Exit Points} 
\label{sec:exitpoints}

The collision axis LOS is used for estimating the exit points of the neutrino beam, and comes from a theoretical model of the layout of the LHC. 
The LOS on either side of the three IPs considered (IP1, IP5 and IP8, hosting the ATLAS~\cite{ATLAS:2008}, CMS~\cite{Chatrchyan:2008zzk}, and LHCb~\cite{LHCb:2008} experiments, respectively) are extrapolated to their exit points from the Earth's surface or above the lakebed in Lake Geneva. IP2, the ALICE~\cite{ALICE:2008} experiment collision point, is not considered since it operates at a much lower luminosity than the other IPs. The luminosity at IP8 is still undecided, but expected to be 10–20\% of that at IP1 and IP5. \medskip 

\begin{figure*}[t!]
\centering
\includegraphics[width=0.9\textwidth]{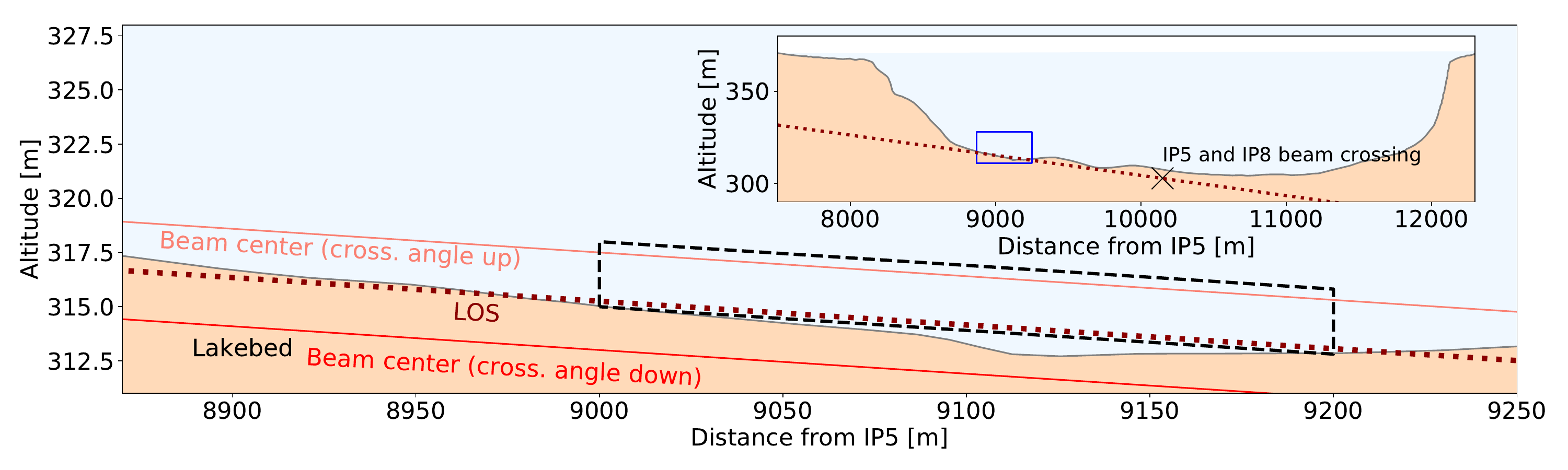}
\caption{The LOS (brown dotted) is estimated to be visible above the lakebed (blue) at a location 9~km away from IP5, providing a possible location for a 200~m long detector (black dashed). The up (light red) and down (red) variations of the crossing angle at IP5 will raise or lower the center of the neutrino beam by 2.25~m. The subplot illustrates the entire lakebed profile along the direction of the LOS, and the point where the nominal LOS from IP5 and IP8 cross is indicated as a cross. Note that uncertainties in the lake depth estimate are not included in the graphic, as the proposed location will in any case have the neutrino beam as high above or as close as possible to the lakebed.}
\label{fig:IP5L_profile}
\end{figure*}

\begin{figure*}
\centering
\includegraphics[width=0.9\textwidth]{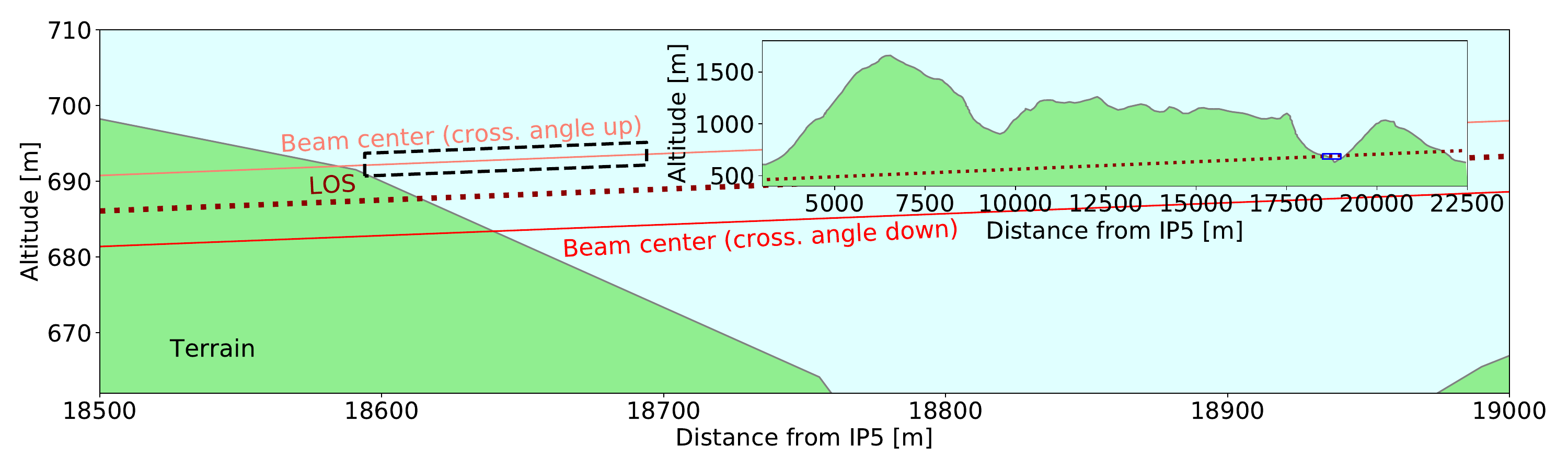}
\caption{To the west of IP5, the LOS (brown dotted) first emerges in a valley in the Jura mountains at a location 19~km away from IP5, providing a possible location for a 10~m long detector (black dashed). The up (light red) and down (red) variations of the crossing angle at IP5 will raise or lower the center of the neutrino beam by 4.7~m. The zoomed-in region in the valley shown as the main plot is indicated as a blue box in the subplot depicting the mountain range profile. The subplot also indicates a further exit point approximately 22~km away from IP5. However, the present work focuses on the closer location to maximize event rates. Note that uncertainties in the terrain height are not included in the graphic, as they could be compensated for by a suitable relocation of the detector up or down the slope after experimentally confirming the path of the beam.}
\label{fig:IP5W_profile}
\end{figure*}

For the extrapolation of the LOS, variations in the terrain are taken into account using digital terrain models: RGE ALTI in France~\cite{RGE_ALTI}, SwissALTI3D in Switzerland~\cite{Swiss_ALTI3D} and Tinitaly in Italy~\cite{Tinitaly}, and the swissBATHY3D~\cite{SwissBathy3D} model is used to get the information of the depth of Lake Geneva.  These models have a resolution of 10~m in Italy, 5~m in France and 2~m in Switzerland (including the depth of Lake Geneva).  The global accuracy of these models is declared as 3.5~m in Italy and better than 1~m in Switzerland and France. Due to the nature of the models, the uncertainty is larger in mountainous areas.  The uncertainty of the location where the LOS reaches the surface of the earth, described below, is also impacted by the rigorous definition of French, Swiss, and Italian geodetic and vertical reference frames (with an error lower than 1~m) and by the uncertainty on the exact location and orientation of the IPs in the LHC. The last point can have a significant impact, as illustrated in \cref{fig:IP5L_profile} and \cref{fig:IP5W_profile}. Overall, the accuracy of the location where the LOS reaches the surface is estimated to be within a few meters, and in some areas possibly a few tens of meters. \medskip

The location of where the LOS reaches the surface of the Earth is shown in~\cref{fig:exitPointMap} for the considered IPs. Due to the slope of the LHC tunnel, and the mountain regions of the Jura and the Alps, the distance between the IP and the exit points vary a lot with the closest exit point on land being 18~km west from the IP5 (denoted IP5W) and the furthest is 183~km from the IP (IP1E). Since the neutrino flux scales inversely as the distance squared, the closest points are the most interesting from a physics point of view.  \cref{fig:exitPointMapZoom} shows the same map, concentrating on the exit points closer than 30~km from their respective IP. In this figure it can also be seen that the IP5L LOS is above the lakebed of Lake Geneva at a distance of about 9~km from the IP for a region of about 200~m. \cref{table:exitPointTable} shows some relevant details of the seven exit points, and compares the  neutrino flux relative to that for IP5L taking into account the distance from the IP and the expected luminosity. This highlights that IP5W and IP5L are the most interesting points to consider further. For comparison the table also shows the information for the locations of the current FASER and SND@LHC experiments, and the proposed FPF. For the exit points in the mountains, it is often the case that the LOS is not parallel to the ground after the exit point, which makes it challenging to place a deep detector (as required for the physics case) at these locations. \medskip

\cref{fig:IP5L_profile} shows the trajectory of the IP5L LOS with respect to the lakebed in the region around 9~km from the IP, and~\cref{fig:IP5W_profile} shows the trajectory for the IP5W LOS with respect to the surface of the Earth over the region 18.5 - 19~km from the IP. In both of these figures the LOS, shown as a brown dotted line, assumes no crossing angle at IP5. In reality there is a small crossing angle at the IPs, pushing the beams vertically or horizontally to avoid additional collisions in the LHC. For the HL-LHC the half crossing angle at IP5 is expected to be 250~$\mu$rad in the vertical plane~\cite{ATSnote}~\footnote{Although a vertical crossing angle in IP5 is the baseline scenario for the full HL-LHC, it is possible that this could be changed to a horizontal crossing for the last part of HL-LHC running to distribute radiation more evenly around the focusing magnets. In this article we assume a vertical crossing for the full HL-LHC program.} moving the LOS by about 2.25~m at 9~km and about 4.75~m at 19~km either up or down. About half of the data is expected to be taken in each configuration. The solid lines in~\cref{fig:IP5L_profile} and~\cref{fig:IP5W_profile} indicate more realistic beam path estimates accounting for the effect of the two crossing angle orientations.

The final focusing of the LHC beam at the collision points introduces a small per-event transverse momentum to the collision system which is quantified by the beam divergence. 
For the expected HL-LHC configuration the divergence will be negligible compared to the intrinsic spread of the neutrino beam from forward hadron production and is therefore not considered in the estimates in this paper.

\section{Detector Designs}
\label{sec:detectors}

While forward neutrino experiments at the LHC are expected to cover a large depth in the beam direction for enhanced total interaction probabilities, the large distances of the considered experiments from IP5 necessitates that they also cover a large transverse area due to the spread of the incoming neutrino flux over distances of several kilometers. Therefore, the experiments at surface level exit points are expected to significantly exceed the size of the existing LHC forward neutrino experiments such as FASER$\nu$~\cite{FASER:2019dxq, FASER:2020gpr} and require us to consider large-volume but cost efficient technologies. 

In the following, we consider three possible detectors at different locations: (i) a water Cherenkov detector placed at the bottom of Lake Geneva, (ii) an emulsion detector submerged in Lake Geneva, and (iii) a kiloton electronic detector in the Jura mountains. In the following, the hypothetical locations of the detectors are provided in further detail, along with basic conceptual designs.

\subsection{FLOUNDER: Water Cherenkov Detector in Lake Geneva}

\begin{figure}[tbh]
\centering
\includegraphics[width=0.35\textwidth,trim={0mm 7mm 0mm 6mm},clip]{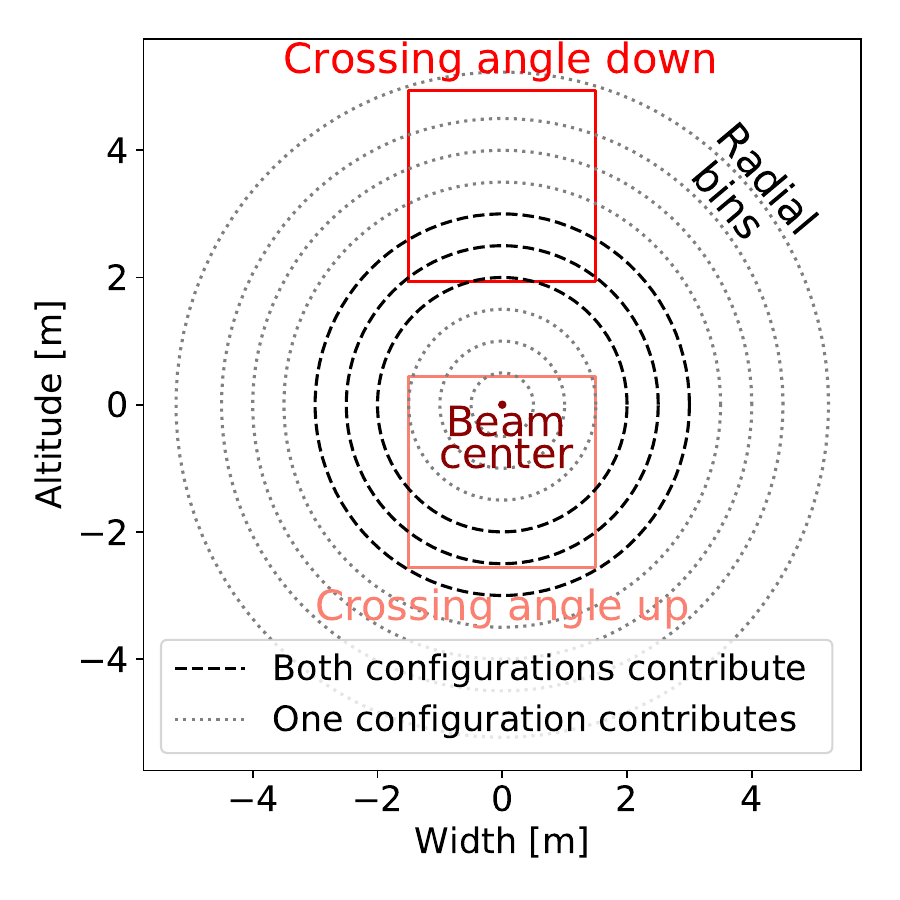}
\caption{The location of the FLOUNDER transverse plane with respect to the center of the neutrino beam (brown dot) when the beam is lowered (red) or raised (light red) by the variations of the crossing angle at IP5. The dashed black circles illustrate radial ranges receiving contributions in both configurations, while the ranges denoted by gray circles are only useful for a single configuration.}
\label{fig:crossingAngle_radialBins}
\end{figure}

The first detector is considered to be placed at IP5L, where the LOS passes through the bottom of Lake Geneva. At this location, the crossing angle effect amounts to a shift of $\pm2.25$~m from the LOS. Notably, a downward shift moves the center of the neutrino beam below the lakebed, as shown in \cref{fig:IP5L_profile}. To increase the expected statistics and physics potential, the detector should cover the center of the beam in the upper crossing angle configuration, while also being close to the lower crossing angle case, enabling it to measure a larger range of rapidities than its nominal size would suggest. To achieve this, we consider a 200~m long benchmark detector with a transverse area of $3 \times 3$~m$^2$, referred to as the Forward LHC Observatory Underwater for Neutrinos and the Dark sEctoR (FLOUNDER). The detector volume is assumed to rest at the bottom of the lake and is outlined in \cref{fig:IP5L_profile} as black dashed lines, while \cref{fig:crossingAngle_radialBins} depicts the radial distances around the neutrino beam center probed in each IP5 crossing angle configuration. The upper crossing angle configuration probes the radial range from 0~m to 3~m, while the lower crossing angle probes the rates of neutrinos from approximately 2~m to 5.2~m, bringing the final pseudorapidity\footnote{Pseudorapidity $\eta \equiv -\log(\tan(\theta/2))$, with $\theta$ the polar angle with respect to the beam line.} coverage of FLOUNDER to $\eta>8.2$. Each crossing angle is assumed to be used for a data collecting  period corresponding to 1.5~ab$^{-1}$, totaling 3~ab$^{-1}$. It should be noted that the above is assuming the best estimate of the LOS position from the study. The uncertainty on the depth of the lake is quoted at 2~m which means the detector could be less optimally located with respect to the LOS by this distance. It would therefore be important to reduce this uncertainty, including characterizing possible time variation in depth, to allow more reliable studies of the physics potential.  

A neutrino beam from CERN provides several interesting design opportunities for a water detector. First, the beam remains tightly collimated at the lake exit point from IP5. The modest cross sectional area of $9~\m^2$ considered here can be inserted within standard industrial cylindrical housing structures such as plastic pipes or non-rusting silo materials. Second, distinct from water or ice Cherenkov detectors searching for astrophysical sources, the neutrinos arrive at fixed times, helping with reduction of background events occurring at random times, and direction, so only one detector dimension requires significant extension. 
This motivates considering an elongated rectangular volume with a strawman design, maximizing the number of neutrino interactions contained within the detector and consistent with the limitations of the topography of the lake bottom, with a scintillator-based system at the front for vetoing or tagging incoming charged particles. Similarly, instrumenting the back end with a scintillator wall would help distinguish final states with one and two muons. A centimeter-level spatial resolution, achievable e.g. with scintillator strips and pixelated photo sensors, would suffice for muons with energies $\gtrsim 20$~GeV, typically traveling 100~m distances.

\begin{table*}[ht!]
\setlength{\tabcolsep}{12pt}
\caption{Comparison of the capabilities of various neutrino detectors and estimates for their systematic uncertainties. Checkmarks (crosses) indicate measurements that we (do not) expect the detector to be able to perform. The $2\mu$ indicates possibility of charm identification via a 2-muon signature, expected to be distinct from single muon final states. Question marks indicate that it is currently unclear how well a detector would perform the task: it is uncertain how well the electron and hadronic final states can be distinguished at FLOUNDER, and if FLARE could identify taus, or charm without two muons.}
\label{table:detectorCapabilities}
\begin{tabular}{l c c c c}
\hline
\hline
~
    & \multicolumn{2}{c}{underground detector} 
    & \multicolumn{2}{c}{surface detector} \\  
    & FASER$\nu$
    & FLARE
    & LED
    & FLOUNDER \\
\hline 
Technology
    & emulsion
    & LAr TPC
    & emulsion
    & water Cherenkov \\
\hline
Electron identification
    & \checkmark
    & \checkmark
    & \checkmark
    & ? \\ 
Muon identification
    & \checkmark
    & \checkmark
    & \checkmark
    & \checkmark \\ 
Tau identification
    & \checkmark
    & ?
    & \checkmark
    & \texttimes \\ 
Charm identification  
    & \checkmark
    & ?, 2$\mu$
    & \checkmark
    & 2$\mu$ \\
\hline
Charge identification
    & $\mu$ ($\tau$)
    & $\mu$ ($\tau$)
    & \texttimes 
    & \texttimes \\ 
Muon momentum resolution
    & $<$20\%
    & $<$5\% 
    & $<$20\%
    & 30\% \\
Muon angle resolution
    & 0.06 mrad
    & $\lesssim$ mrad
    & 0.06 mrad
    & 5 mrad\\
$E_{\rm had}$ resolution
    & 30\%
    & 30\%
    & 30\%
    & $\gtrsim$30\% \\
\hline
$\nu_e$ energy resolution
    & 30\%
    & 20\%
    & 30\%
    & $\gtrsim$30\% \\
Transverse position resolution
    & $\sim$ $\mu$m
    & $\sim$ mm
    & $\sim$ $\mu$m
    & $\sim$ 10~cm \\
Longitudinal position resolution
    & $<$ mm
    & $\sim$ mm
    & $<$ mm
    & $\sim$ 1~m \\ 
\hline
Flux (relative to FASER$\nu$)
    & 1
    & 0.6
    & 0.002
    & 0.002 \\ 
Target mass (tons)
    & 1
    & 10
    & 200
    & 7500 \\ 
Event rate per luminosity (relative to FASER$\nu$)
    & 1
    & 12
    & 0.8
    & 15 \\ 
\hline
\hline
\end{tabular}
\end{table*}

Nonetheless, there are several challenges to consider. The range of neutrino energies produced by proton collisions is rather broad, extending roughly from 10 GeV to a few TeV, and the energy of any specific neutrino is a priori unknown.  This is due to the fact that the events which produce the highest-energy neutrinos are typically not within the acceptance or triggered by the central experiment at the IP. Nor will the flavor of the neutrino entering FLOUNDER be known. Further, a water Cherenkov detector buried under less than 60~m of lake water will require a sealed volume to eliminate extraneous light from the sun and other sources. Though the attenuation length requirements are likely to be satisfied with clean lake water, prefiltering is desirable to maintain uniformity over time and to minimize bio-fouling of the detector surfaces that observe the Cherenkov light emission from neutrino interactions. If the interior volume is initially filled with purified water, coupled with the expected dark conditions within the sealed volume, the rate of bio-fouling may be acceptable without the need for continuous filtering. We note that neither the HAWC water tanks nor the Pierre Auger Observatory water tanks require continuous water purification~\cite{PierreAuger:2007kus, historical:2023opo}. 

At the proposed scale of the experiment, the spatial granularity of the photosensor coverage could easily be made greater than in typical water Cherenkov detectors designed to observe neutrinos from astrophysical sources. On the other hand, the majority of neutrino energies from the LHC exceed 50 GeV, much higher than in most water Cherenkov detectors designed for accelerators. This implies that less fractional photocathode area is required, and the detector can be relatively sparsely populated with photo-collectors, nominally assumed to be photomultiplier tubes (PMT) with a minimum diameter of 8~cm. A strawman design with 4 PMTs per square meter should provide $>$40 photoelectrons per GeV of deposited energy. About 10,000 PMTs are required to cover the interior walls of the strawman detector, a large but manageable number. The PMTs will measure the arrival time distribution of the Cherenkov photons with a precision of $\sim$ 1 ns.  The spatial pattern of PMTs and their timing distributions provides information to reconstruct the development and evolution of showers and the path of long-range charged particles emerging from the showers. 
 
A uniform placement of PMTs on the walls of FLOUNDER yields accurate transverse coordinate information but worse longitudinal accuracy. For instance, a transverse (longitudinal) vertex resolution of $<0.5$~m ($< 5$~m) is estimated for the KM3NeT experiment~\cite{KM3Net:2016zxf}. Also the Hyper-K experiment is expected to have a vertex resolution of $<0.5$~m~\cite{Hyper-Kamiokande:2018ofw}. Due to the similarity of the detector technologies, these can be taken as benchmarks for FLOUNDER estimates. Optimistically however, the smaller track-to-PMT distance at FLOUNDER can improve on this resolution, but remains in the same order of magnitude. Hence, a transverse (longitudinal) resolution of $\sim 10$~cm ($\sim 1$~m) is assumed. Neither of these assumptions strongly impact the science reach of FLOUNDER.

For KM3NeT, a 29\% muon momentum resolution and at worst a median 0.5$^\circ = 8.7$~mrad angle difference between reconstructed and simulated tracks are estimated for energies at the TeV scale~\cite{Drakopoulou:2016azl}, while the KM3NeT 2 letter of intent states a 0.3-0.4$^\circ$ angular uncertainty at a few TeV~\cite{KM3Net:2016zxf}. 
However, the KM3NeT photomultiplier tube strings are spaced an order of magnitude further apart than the half-diameter of FLOUNDER, and a slightly improved muon angle resolution of 5~mrad can be estimated at FLOUNDER. 
The strawman design proposed for FLOUNDER provides an increased opportunity to measure stochastic light depositions by high energy muons emerging from the charged-current (CC) $\nu_{\mu}$ vertex. Such a detector should also be able to measure the energies of high energy muons in a large volume with a reasonable precision, $\sim30\%$. In contrast, measuring them with a spectrometer over a large volume would be considerably more challenging experimentally, requiring expensive technologies. 
IceCube has found that starting track events (the subsample of events where the interaction occurs within a fiducial volume of the detector, comparable to the event geometry of FLOUNDER) can be reconstructed with an energy resolution of $25-30\%$ for energies between 1 TeV and 10 PeV~\cite{Silva:2023wol}.  Given the much higher granularity of FLOUNDER than IceCube, an optimistic value of 30\% is assumed at FLOUNDER down to neutrino energies of 10 GeV. The same resolution is optimistically assumed for the reconstruction of the energies of electron neutrinos and hadronic recoils.
\cref{table:detectorCapabilities} summarizes the assumed experimental capabilities and systematic uncertainty estimates of the FLOUNDER detector are summarized and compared with the FASER$\nu$ and FLARE~\cite{Anchordoqui:2021ghd, Feng:2022inv, Batell:2021blf} detectors.

\subsection{Lake Emulsion Detector}

The building of a large underwater detector at the bottom of Lake Geneva involves significant engineering challenges. A likely first step towards the realization of such a project would be a relatively small proof-of-principle apparatus to verify the expected event rate. Similar to the initial investigation of the underground near locations, this could be performed using an emulsion detector~\cite{FASER:2021mtu}. 
However, instead of serving merely as a pathfinder experiment, such a detector could have individual merit as a larger long-term experiment complementary to FLOUNDER due to the very different experimental capabilities of the technology. 
We consider the prospect of submerging a detector emulating the construction of FASER$\nu$(2)~\cite{FASER:2019dxq, FASER:2020gpr}, consisting of heavy metal, e.g. tungsten, plates interleaved with emulsion films, at the proposed location, for instance inside a shipping container. Here, this setup is referred to as the lake emulsion detector (LED). A portion of the neutrinos passing through the detector will interact in the tungsten, and are identified by the tracks of the final state charged particles in the emulsion films. A clear advantage of such a detector is that it does not require services, such as electricity, thereby significantly reducing the associated cost. 

Unlike the FASER and FPF caverns, the IP5L location does not suffer from a large flux of highly energetic muons. Muons produced in hadron decays close to the IP are stopped in the roughly $9~\km$ of rock before reaching the lake. The remaining flux of muons originates mainly from neutrino interactions in the last few kilometers of rock as well as cosmic rays. In comparison to the underground near locations, this would significantly reduce the necessary frequency of collecting and replacing the emulsion films. In addition, a smaller muon flux also reduces the complexity of event reconstruction. This would allow for a reduced number of emulsion films by increasing the thickness of the target plates from $1~\mm$ currently used at FASER to about $1~\cm$, leading to the realization of a 10 times heavier detector for a similar cost.

As a first result, this design allows to compare the observed event counts to theoretical expectations and to validate the case for further lake detector development. In addition, such a detector provides energy resolution and lepton identification capabilities sufficient for physics analyses~\cite{FASER:2024hoe}. Electrons can be identified via their dense electromagnetic shower while muons are long tracks that traverse many nuclear interaction lengths of material. In contrast to the other considered detector technologies, an emulsion detector can also reliably observe tau neutrinos through the identification of the decay vertex of the tau lepton. This opens the opportunity to study tau neutrino interactions, measure $\nu_\tau$ CC cross sections, and use these to test lepton flavor universality in the neutrino sector. The measurement capabilities of such an emulsion detector, including energy and spatial resolution, are summarized in \cref{table:detectorCapabilities}, where we adapt the performance estimates presented for FASER in Ref.~\cite{FASER:2019dxq}.

When estimating the expected event rate of the LED, we first assume a FASER$\nu$2-size detector, consisting of a 20~ton tungsten target with dimensions $0.4 \times 0.4 \times 6.6$~m$^3$, that is centered around the LOS at the IP5L location at the bottom of Lake Geneva. We note again that, due to the crossing angle, the LOS is above the lakebed only for about half the time and therefore we assume that the detector is exposed to $1.5~\text{ab}^{-1}$. Since the reduced muon flux may permit thicker tungsten plates and therefore a heavier detector, we also present results for a 200~ton tungsten target with dimensions $1.2 \times 1.2 \times 7.3$~m$^3$.

\subsection{Electronic Detector in Jura Mountains}

Another promising location for a surface detector is IP5W, located approximately 19 km from the interaction point in the Jura mountains, as shown in Figure \ref{fig:IP5W_profile}. A site visit showed that the LOS emerges from the surface in a field located on a sloped hill in a valley, accessible by car. The slope of the mountain surface is calculated to be 0.166~rad downward, while the beam slope is 0.014~rad upward. The crossing angle introduces vertical shifts of $\pm$4.75 m in the beam position, which corresponds to a shift of the emerging point by $\mp$26 m in the beam direction. 

\begin{table*}
\setlength{\tabcolsep}{4pt}
\caption{Event rates based on the statistics of neutrino flavors coming from IP5. Both the up and down variations of the IP5 crossing angle contribute significantly to the rates at FLOUNDER. The LED and the IP5W detectors are assumed to be centered at the upward crossing angle LOS. The contributions during the downward variation are thus negligible and the luminosity is effectively 1.5~ab$^{-1}$.}
\label{table:eventRates}
\begin{tabular}{l c c c c c c c c c c c c}
\hline
\hline
~ & $\mathcal{L}$
  & distance
  & Dimensions
  & Volume
  & $M_{\rm Target}$
  & Rapidity
  & \multicolumn{2}{c}{$\nu_e$} 
  & \multicolumn{2}{c}{$\nu_\mu$} 
  & \multicolumn{2}{c}{$\nu_\tau$} \\
~ & ab$^{-1}$  
  & km
  & $\text{m} \times \text{m} \times \text{m}$
  & m$^3$
  & ton
  & ~
  & CC
  & NC
  & CC
  & NC
  & CC
  & NC\\
\hline
FASER$\nu$ Run3
    & 0.25
    & 0.48
    & $0.25 \times 0.25 \times 1.0$
    & 0.063
    & 1.1
    & $>8.9$
    & 1.9k
    & 590
    & 9.2k
    & 2.9k
    & 34
    & 12\\
FASER$\nu$ HL
    & 3
    & 0.48
    & $0.25 \times 0.25 \times 1.0$
    & 0.063
    & 1.1
    & $>8.9$
    & 22k
    & 7.1k
    & 110k
    & 34k
    & 410
    & 140\\
FASER$\nu$2
    & 3
    & 0.62
    & $0.4 \times 0.4 \times 6.6$
    & 1.1
    & 20
    & $>8.7$
    & 220k
    & 69k
    & 1.1M
    & 340k
    & 4.3k
    & 1.5k\\
\hline
LED-20T
    & 1.5
    & 9.0
    & $0.4 \times 0.4 \times 6.6$
    & 1.1
    & 20
    & $>11$
    & 680
    & 220
    & 4.0k
    & 1.2k
    & 11
    & 3.7\\
LED-200T
    & 1.5
    & 9.0
    & $1.2 \times 1.2 \times 7.3$
    & 11
    & 200
    & $>10.3$
    & 7.6k
    & 2.4k
    & 39k
    & 12k
    & 110
    & 37\\
\hline
IP5W (NuTeV)
    & 3
    & 19
    & $3 \times 3 \times 10$
    & 90
    & 690
    & $>9.4$
    & 12k
    & 3.8k
    & 60k
    & 19k
    & 170
    & 58\\
IP5W (NOvA)
    & 3
    & 19
    & $15 \times 15 \times 60$
    & 13500
    & 15000
    & $>8.5$
    & 130k
    & 44k
    & 650k
    & 210k
    & 2.9k
    & 1.0k\\
\hline
FLOUNDER
    & 3
    & 9.0
    & $3 \times 3 \times 200$
    & 1800
    & 1800
    & $>8.2$
    & 78k
    & 25k
    & 380k
    & 120k
    & 1.6k 
    & 590\\ 
Cross.angle $\uparrow$
    & 1.5
    & ~
    & ~
    & ~
    & ~
    & $>8.9$
    & 49k
    & 16k
    & 250k
    & 81k
    & 890
    & 320\\ 
Cross.angle $\downarrow$
    & 1.5
    & ~
    & ~
    & ~
    & ~
    & 8.2 -- 9.1
    & 29k
    & 9.4k
    & 130k
    & 43k
    & 760
    & 270\\ 
\hline
\hline
\end{tabular}
\end{table*}

\begin{figure*}[t!]
\centering
\includegraphics[width=0.92\textwidth,trim={0mm 0mm 0mm 2mm},clip]{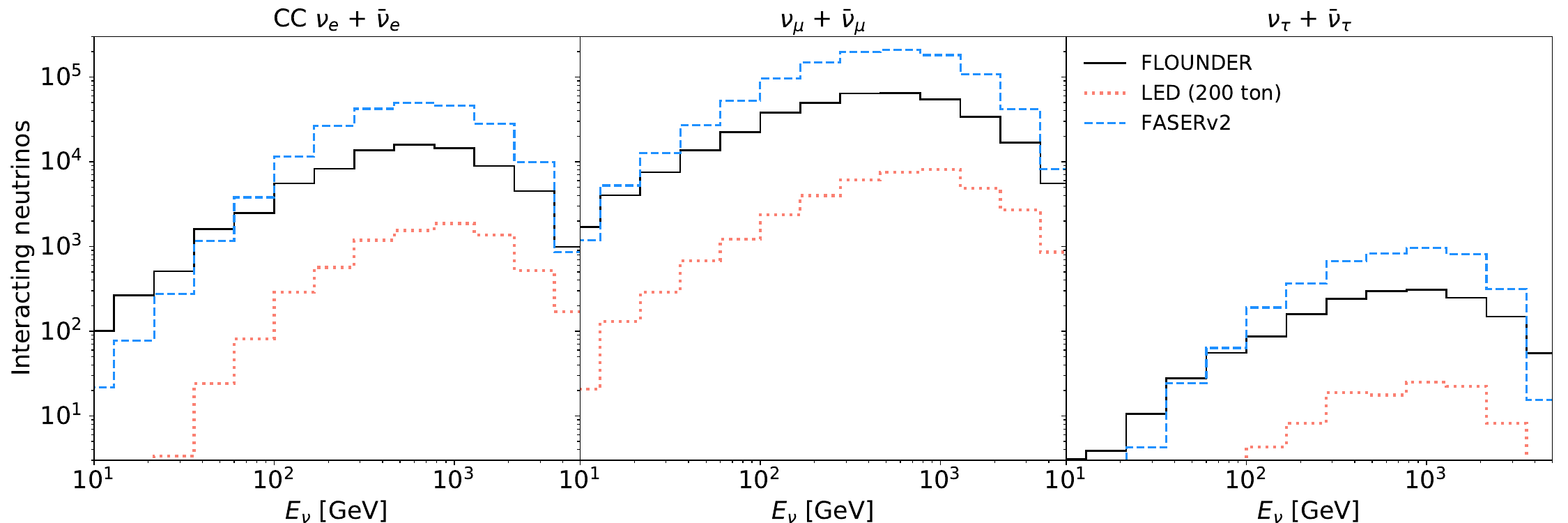}
\caption{A comparison of the theoretical expectation for the number of CC interactions as a function of neutrino energy at various detectors. FASER$\nu$2 is assumed to be located at the proposed FPF location, while LED represents the rates obtainable with a 200~ton emulsion detector located in the lake at 9~km away from the IP, corresponding to the FLOUNDER location.}
\label{fig:detectorComparison}
\end{figure*}

The above conditions allow the placement of a large-volume electronic neutrino detector on the surface. Historically, several neutrino experiments existed to detect multi-100 GeV accelerator neutrinos, such as CDHSW~\cite{Holder:1977gn, Holder:1977em}, CHARM~\cite{CHARM:1980ppk}, CCFR and NuTeV~\cite{Bolton:1990si}, where the latter had a target mass of 690~tons. The expected neutrino event rates for a NuTeV-sized detector are shown in \cref{table:eventRates}. We find that it is about half of the rate to be collected with a 1~ton detector at the FASER location during the HL-LHC. This indicates that such a detector would not provide sensitivity beyond those in the existing near detector locations. 

Even larger detectors were built for long-baseline neutrino oscillation measurements, including OPERA~\cite{OPERA:1999hzg}, MINOS~\cite{Lang:2001rw} and NOvA~\cite{NOvA:2007rmc}. For illustration, the event rates are also estimated for a NOvA-sized detector with roughly a 14~kiloton mass placed at the IP5W location. As presented in \cref{table:eventRates}, it collects about six times the statistics of a 1~ton detector operating during the HL-LHC era at the FASER location, and about two-thirds of the statistics of FASER$\nu$2. While this illustrates the mass scale needed for an IP5W surface detector to surpass the planned upgrades of FASER and SND@LHC, we will not consider it in the remainder of this article. 

\section{Neutrino spectra}
\label{sec:spectra}

The neutrinos emerging from the primary proton collision at IP5 are mainly produced through hadron decays, with decays from pions, kaons and charm mesons dominating and giving different proportions to the flux of each neutrino flavor. We simulate the production of  light hadrons using \texttt{EPOS-LHC}~\cite{Pierog:2013ria} and generate the charm meson spectra using \texttt{POWHEG}~\cite{Nason:2004rx, Frixione:2007vw, Alioli:2010xd} matched with \texttt{Pythia~8.3}~\cite{Bierlich:2022pfr} for parton shower and hadronization, as described in Ref.~\cite{Buonocore:2023kna}. Light mesons are long-lived and decay downstream of the interaction point. To propagate mesons through the LHC's beam pipe and magnetic fields as well as to simulate their decays, we use the neutrino flux simulation introduced in Ref.~\cite{Kling:2021gos}, with the HL-LHC configuration described in Ref.~\cite{FASER:2024ykc}. Absorptions of neutrinos in the rock were found to have negligible effects on the beam intensity. To obtain the expected event rates, we use the neutrino interaction cross section provided by \texttt{GENIE}~\cite{Andreopoulos:2009rq}. We note that the Bodek-Yang model~\cite{Bodek:2002vp, Bodek:2004pc, Bodek:2010km} employed in \texttt{GENIE} agrees with more recent cross section calculations for high energy neutrinos~\cite{Candido:2023utz, Jeong:2023hwe} within an uncertainty of $\lesssim 6\%$~\cite{FASER:2024ykc}.

\begin{figure*}[t!]
\centering
\includegraphics[width=0.62\textwidth,trim={0mm 3mm 0mm 3mm},clip]{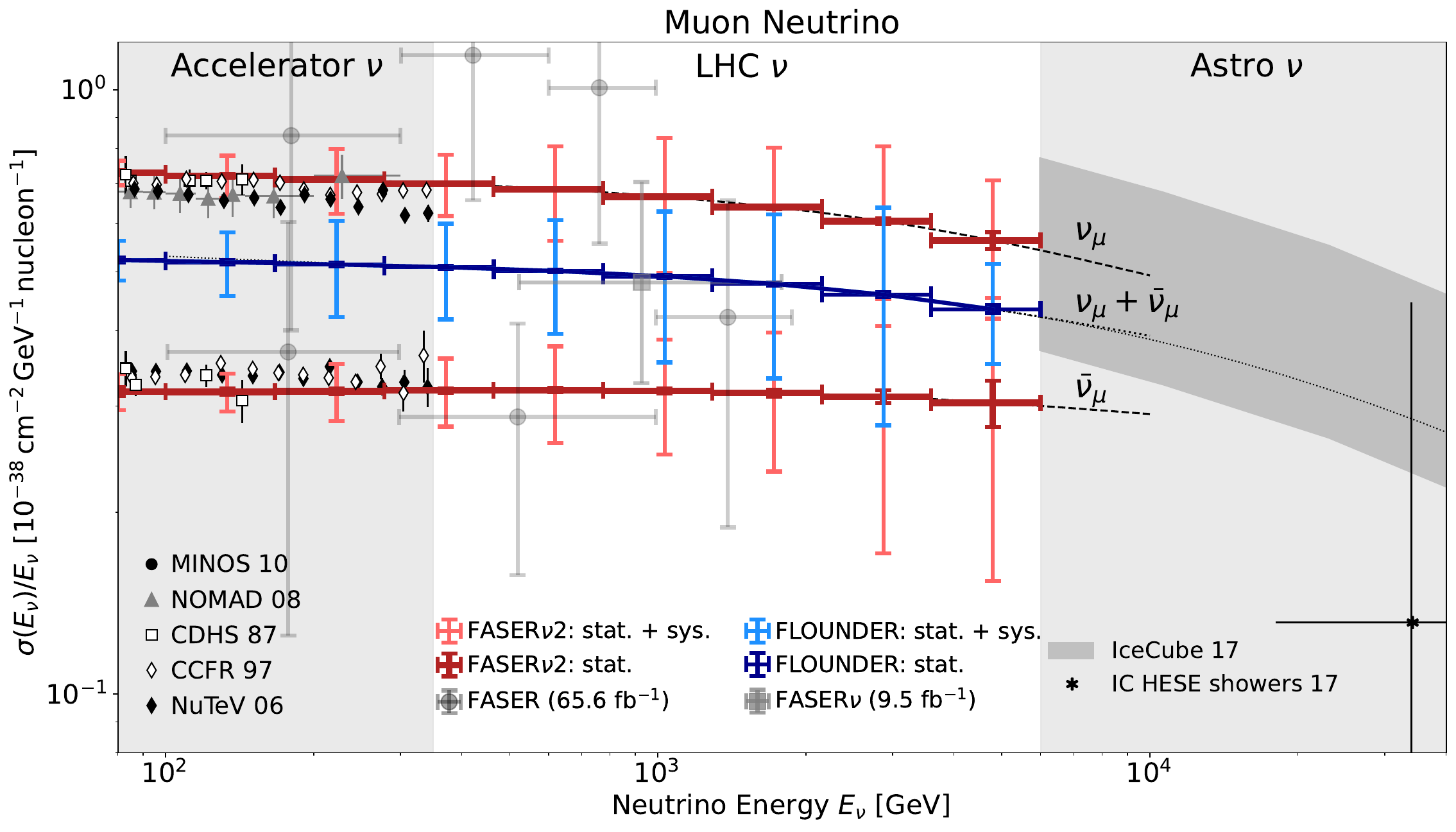}
\includegraphics[width=0.337\textwidth,trim={0mm 3mm 0mm 3mm},clip]{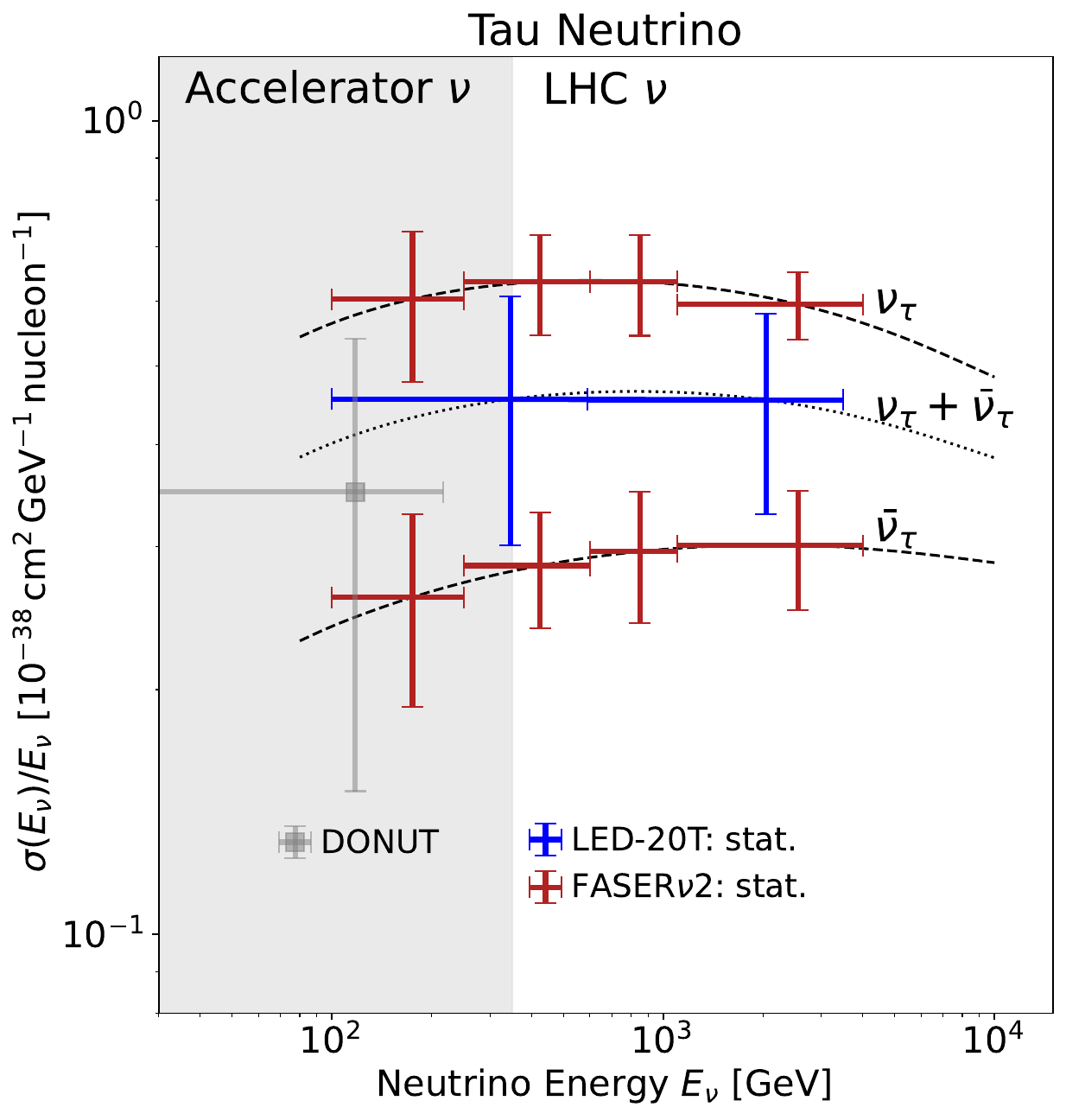}
\caption{Projections of $\nu_\mu$ CC cross section measurements at FLOUNDER (left panel) and $\nu_\tau$ CC cross section measurements at LED (right panel). 
Projections for FASER$\nu$2 (red) at the FPF, and the recent measurements at FASER$\nu$ and FASER (grey), are shown for comparison. Muon neutrino cross section measurements are complemented by cross section measurements at accelerators at lower energies $\lesssim 400$ GeV, and at IceCube at high energies $\gtrsim 6$ TeV while the tau neutrino cross section has only been measured by the DONUT collaboration. For the muon neutrino cross section measurement, we also compare the level of precision reached by considering only statistical errors (dark error bars), and after including neutrino flux uncertainties (light error bars) which are the dominant systematic. To describe the neutrino flux uncertainties, we use the tuning uncertainty following the forward \texttt{Pythia} tune~\cite{Fieg:2023kld} for light hadrons, and use variations in the resummation and factorization scales for charm hadrons, as described in Refs.~\cite{Buonocore:2023kna,FASER:2024ykc}. In both panels, we also plot the cross section prediction for deep inelastic scattering using the Bodek-Yang model in dashed lines, and the average cross section between $\nu$ and $\overline{\nu}$ in the dotted line.}
\label{fig:crossSection}
\end{figure*}

The total numbers of neutral current (NC) and CC interactions in the considered surface detectors, together with operating and proposed underground detectors, are shown in \cref{table:eventRates}. The corresponding spectra for the total number of CC interactions as a function of the incoming neutrino energy $E_\nu$ are illustrated in \cref{fig:detectorComparison}. During the HL-LHC run, corresponding to 3~ab$^{-1}$, a 20~ton emulsion detector submerged in the lake could be expected to yield about a third of the event rates of FASER$\nu$ during Run~3. A 200~ton detector increases the expected event rate tenfold, providing about four times the statistics of FASER$\nu$ or a third of the statistics achieved with a 1.1~ton neutrino detector at the FASER location during the HL-LHC era. In contrast, FLOUNDER yields significantly greater rates than FASER$\nu$ due to its large volume and 1.8~kiloton target mass. Larger detector sizes naturally increase the rates in all cases.

The existing and proposed underground detectors close to the IP are exposed to a large muon flux, with a rate of about 1~Hz/cm$^2$. The situation changes for the considered surface detectors. A minimum ionizing particle loses about 500~GeV/km when passing though rock, while muons with energies above 300~GeV suffer even higher radiative energy losses. Therefore, beam muons will not reach the considered surface detectors. High-energy muons entering the front of the detector and passing through it must therefore originate from muon neutrino interactions in the rock in front of the detector, providing an alternative way to detect and study these neutrinos. 
For this, muon neutrino interactions in a $10~\m\times 10~\m \times 2~\km$ volume in front of FLOUNDER are simulated using \texttt{Pythia}. The muons are propagated accounting for both the average energy loss and deflection through multiple-Coulomb-scattering following the description in Ref.~\cite{ParticleDataGroup:2020ssz}.  Roughly 2~million muons are expected to enter FLOUNDER through the upstream boundary, significantly enhancing the statistics compared to interactions in its target volume. The energy spectrum of the entering muons has an average energy of a few 100~GeV, and is discussed below in \cref{subsec:NeutrinoFluxMeasurements}. Notably however, such a measurement only provides restricted information on the neutrino interaction: the vertex location, the muon energy at the vertex, and the energy of the hadronic recoil system remain inaccessible, limiting the physics applications of such muon data. 

\section{Physics applications} 
\label{sec:applications}

We now turn to the physics case of surface neutrino experiments. The lake emulsion detector is qualitatively and quantitatively akin to the stand-alone FASER$\nu$ detector, and we refer the reader to a discussion of the physics potential in Ref.~\cite{FASER:2019dxq}. Hence the discussion focuses on FLOUNDER, with remarks on emulsion detection when applicable. We consider selected neutrino physics and dark sector benchmark scenarios receiving recurring attention in the context of forward LHC experiments. 

\subsection{Cross Sections at TeV Energies}

\begin{figure*}[t!]
\centering
\includegraphics[width=0.32\textwidth,trim={0mm 2mm 0mm 4mm},clip]{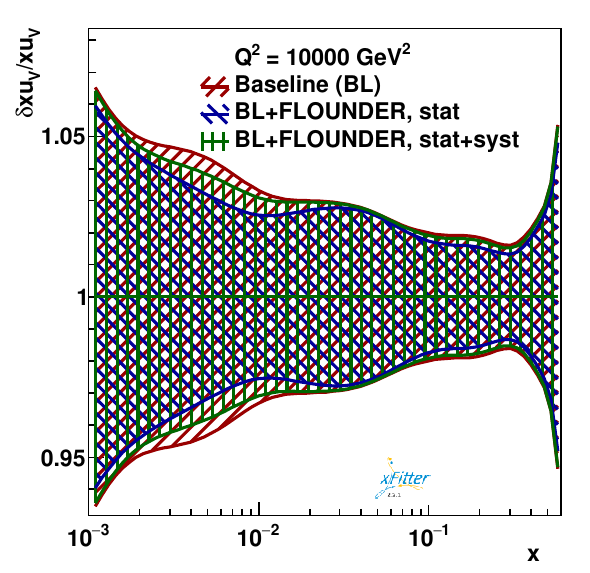}
\includegraphics[width=0.32\textwidth,trim={0mm 2mm 0mm 4mm},clip]{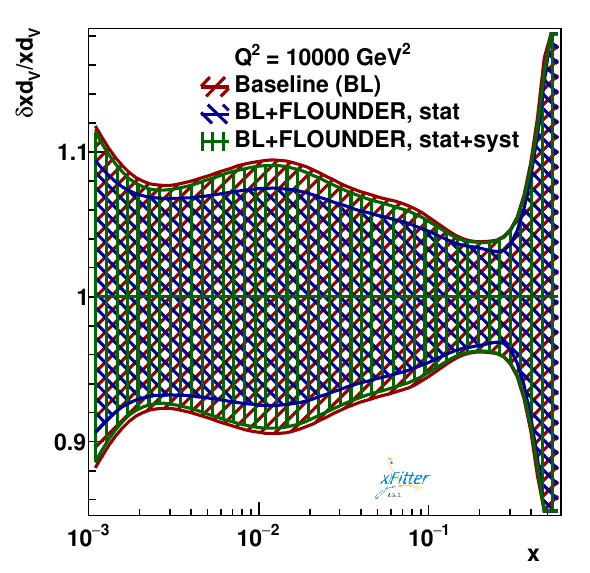}
\includegraphics[width=0.32\textwidth,trim={0mm 2mm 0mm 4mm},clip]{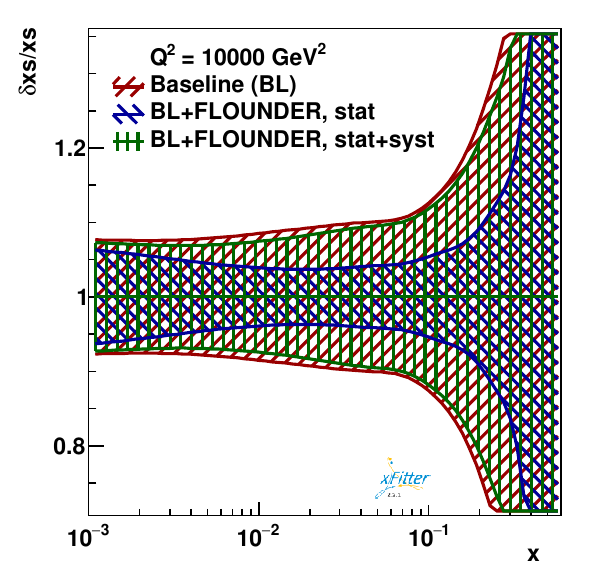}
\caption{The PDF4LHC21 proton PDFs (red) for the up valence quark $u_V$ (left), the down valence quark $d_V$ (center) and the strange quark $s$ (right), compared with the results of PDF profiling with FLOUNDER pseudodata assuming statistical uncertainties only (blue) as well as statistical and approximate systematic uncertainties (green). The optimistic case with only statistical uncertainties yields some improvement, which is however greatly reduced after including systematic uncertainties.}
\label{fig:PDFs}
\end{figure*}

Prior to the recent measurements by the FASER Collaboration~\cite{FASER:2024hoe, FASER:2024ref}, neutrino interaction cross sections have been measured using either low-energy accelerator or very high-energy astrophysical neutrino data. The gap between these regimes is bridged by observing the most energetic artificially produced neutrinos at the LHC, and the sizable FLOUNDER event statistics allow measuring the inclusive muon neutrino interaction cross section. \cref{fig:crossSection} shows the precision reachable for CC cross section measurements at FLOUNDER and FASER$\nu$2 assuming statistical errors, and after including the current neutrino flux uncertainties, which are expected to be the dominant systematic. We find that the measurement is systematics limited for muon neutrinos, which results from the large event rate. In~\cref{fig:crossSection}, we compare these projections to the cross sections measured using accelerator neutrinos at MINOS~\cite{MINOS:2009ugl}, NOMAD~\cite{NOMAD:2007krq}, CDHS~\cite{Berge:1987zw}, CCFR~\cite{Seligman:1997fe} and NuTeV~\cite{NuTeV:2005wsg} as well as astrophysical data at IceCube~\cite{IceCube:2017roe, Bustamante:2017xuy}. We note that additional measurements will also be performed by the existing FASER experiment using its emulsion detector and electronic components~\cite{FASER:2019dxq, Boyd:2882503, Arakawa:2022rmp}. 

Due to the lack of charge identification, and potential difficulties in recognizing electrons, FLOUNDER data is best applicable for measuring the $\nu_\mu + \bar{\nu}_\mu$ cross section (FASER$\nu$2 is able to identify the outgoing muon's charge using the FASER2 magnet). If however $\nu_e$ CC and NC interactions can be separated reliably, FLOUNDER data could, in principle, also be used for measuring the Weinberg angle~\cite{NuTeV:2001whx} at different energies than previously measured, constraining non-standard interactions~\cite{Ismail:2020yqc} and observing the neutrino charge radius~\cite{MammenAbraham:2023psg}.

On the other hand, LED is capable of cross section measurements for electron and tau neutrinos. Notably, the latter has so far only been measured by the DONUT collaboration~\cite{DONuT:2007bsg}. In the right panel of \cref{fig:crossSection} we show projections for the $\nu_{\tau} + \bar{\nu}_{\tau}$ cross section measurement at LED. In contrast to muon neutrinos, tau neutrinos are less copiously produced and are also less collimated, resulting in a significantly smaller event rate at far detectors. Thus, the precision that can be reached for the tau neutrino cross section at LED is statistics limited. We also show projected results for FASER$\nu$2 which can, in combination with the FASER2 spectrometer, measure the $\nu_{\tau}$ and $\bar{\nu}_{\tau}$ cross sections separately utilizing a subset of events in which the tau decays to a muon. 

\subsection{Constraining Proton Structure} 

CC neutrino interactions can be used for testing proton structure, providing complementary information to the deep inelastic scattering data obtained at HERA or fixed target collisions relying on NC interactions. 
Such measurements could reduce overall PDF uncertainties for key LHC measurements, such as Higgs or electroweak boson production~\cite{LHCHiggsCrossSectionWorkingGroup:2016ypw}, and help to break degeneracies between PDF parameters and effective field theory coefficients~\cite{Kassabov:2023hbm}. 
Following the strategy of Ref.~\cite{Cruz-Martinez:2023sdv}, the potential of FLOUNDER to constrain PDFs via $\nu_\mu$ CC interactions is assessed by performing Hessian profiling~\cite{Paukkunen:2014zia,  Schmidt:2018hvu, AbdulKhalek:2018rok, HERAFitterdevelopersTeam:2015cre} implemented into the \textsc{xFitter} open-source QCD analysis framework~\cite{Alekhin:2014irh, Bertone:2017tig, xFitter:2022zjb, xFitter:web}. Assuming a free isoscalar nucleon target
\footnote{Water is not isoscalar, but this is not expected to have a strong influence on the qualitative result.}
, the study is performed using the PDF4LHC21~\cite{PDF4LHCWorkingGroup:2022cjn} proton PDF set, which is a Monte Carlo combination~\cite{Watt:2012tq, Carrazza:2015hva} of the CT18~\cite{Hou:2019efy}, MSHT20~\cite{Bailey:2020ooq}, and NNPDF3.1~\cite{NNPDF:2017mvq} global PDFs with Hessian representations obtained via the methods in Refs.~\cite{Gao:2013bia, Carrazza:2015aoa, Carrazza:2016htc}. 

The resulting fractional PDF uncertainties are shown for the up and down valence quarks, $u_V$ and $d_V$, and the strange quark $s$ in~\cref{fig:PDFs} at $Q^2 = 10 000$~GeV$^2$. The analysis performed assuming only statistical uncertainties indicates modest improvement from the PDF4LHC21 baseline. However, the inclusion of estimated experimental systematic uncertainties renders the profiled PDFs equivalent to the baseline. 

\begin{figure*}[t!]
\centering
\includegraphics[width=0.32\textwidth,trim={0mm 0mm 35mm 0mm},clip]{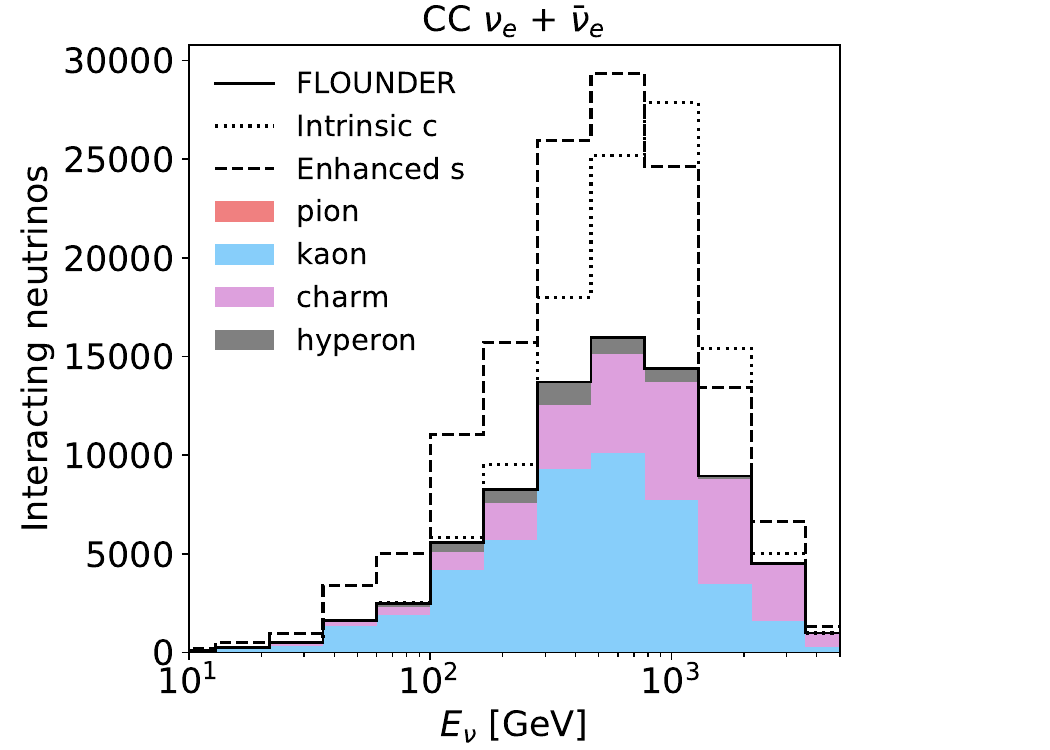}
\includegraphics[width=0.32\textwidth,trim={0mm 0mm 35mm 0mm},clip]{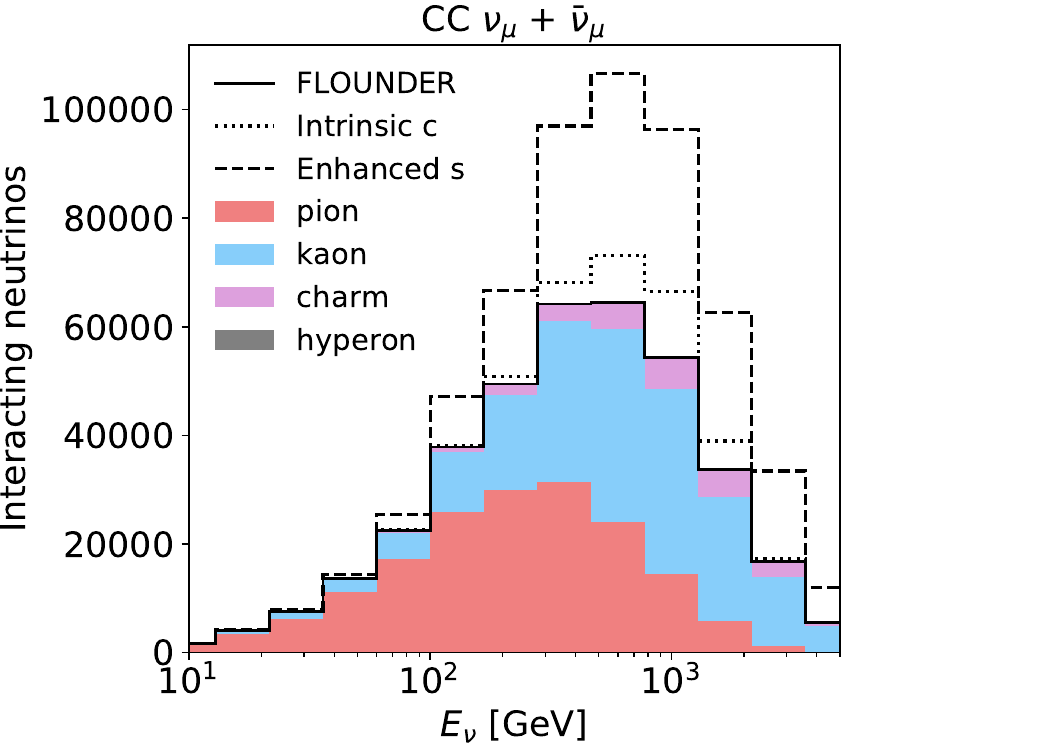}
\includegraphics[width=0.32\textwidth,trim={0mm 0mm 35mm 0mm},clip]{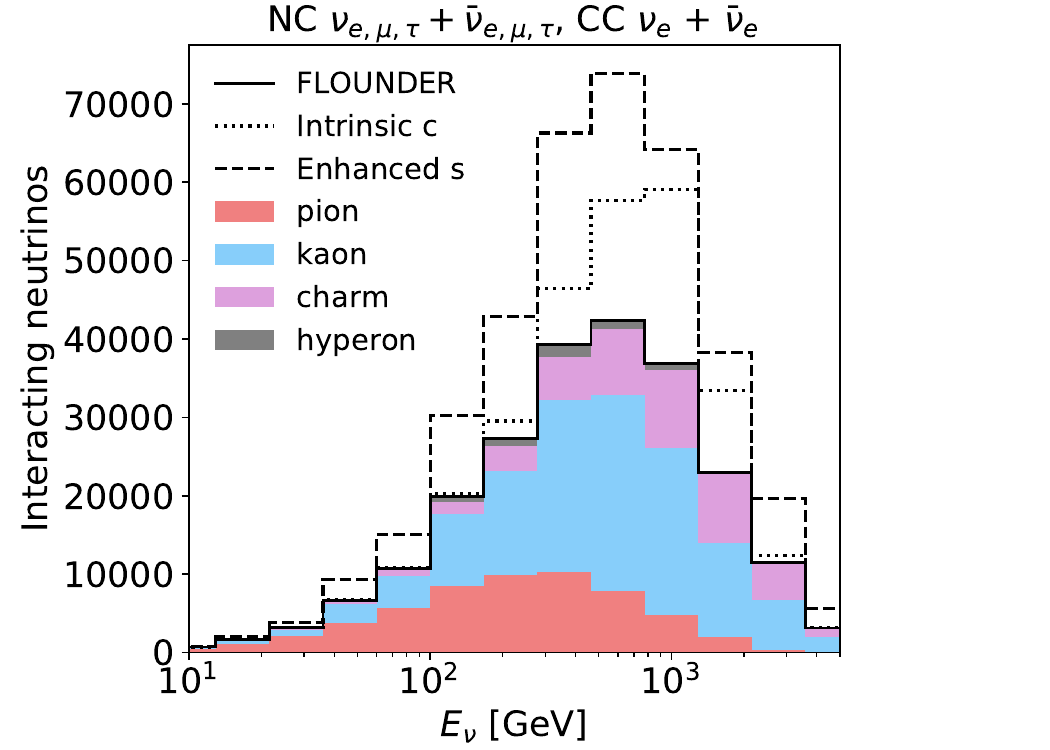}
\caption{The parent hadron composition of the spectra of $\nu_e+\overline{\nu}_e$ (left) and $\nu_\mu+\overline{\nu}_\mu$ (middle) CC neutrino interactions in FLOUNDER as a function of the incoming neutrino energy. The right panel shows the energy spectrum for cascade-like interactions, consisting of both $\nu_e+\overline{\nu}_e$ CC and NC events of all flavors. 
Changes of the energy spectrum caused by forward strangeness enhancement, following Refs.~\cite{Anchordoqui:2022fpn, Sciutto:2023zuz}, and an intrinsic charm component, following the BHPS model~\cite{Brodsky:1980pb} implemented in the CT14 PDF~\cite{Hou:2017khm} as estimated in Ref.~\cite{Maciula:2022lzk}, are shown as dashed and dotted lines, respectively.
See text for details.}
\label{fig:composition_FLOUNDER}
\end{figure*}

\begin{figure*}[t!]
\includegraphics[width=0.32\textwidth,trim={2mm 0mm 35mm 0mm},clip]{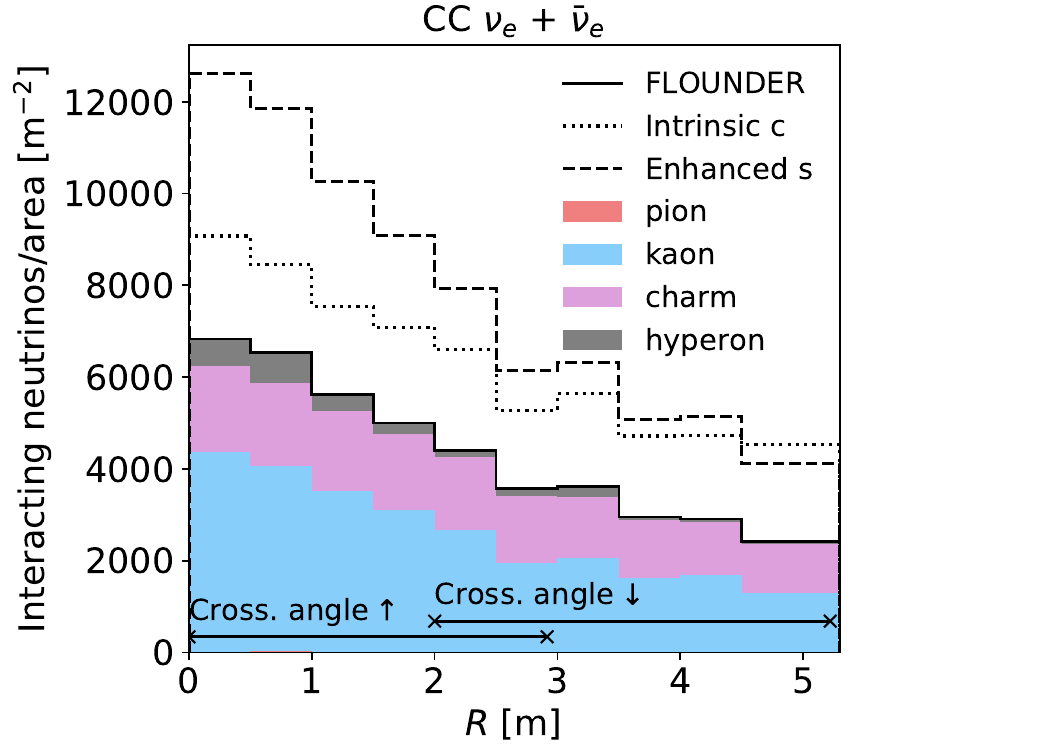}
\includegraphics[width=0.32\textwidth,trim={2mm 0mm 35mm 0mm},clip]{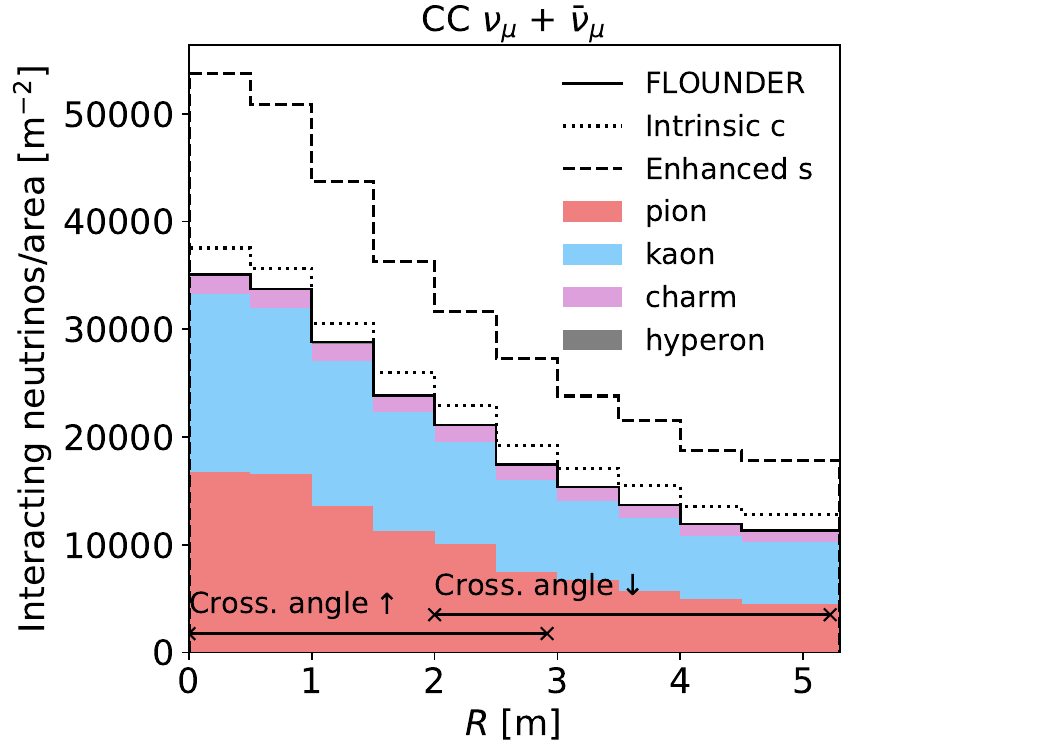}
\includegraphics[width=0.32\textwidth,trim={2mm 0mm 35mm 0mm},clip]{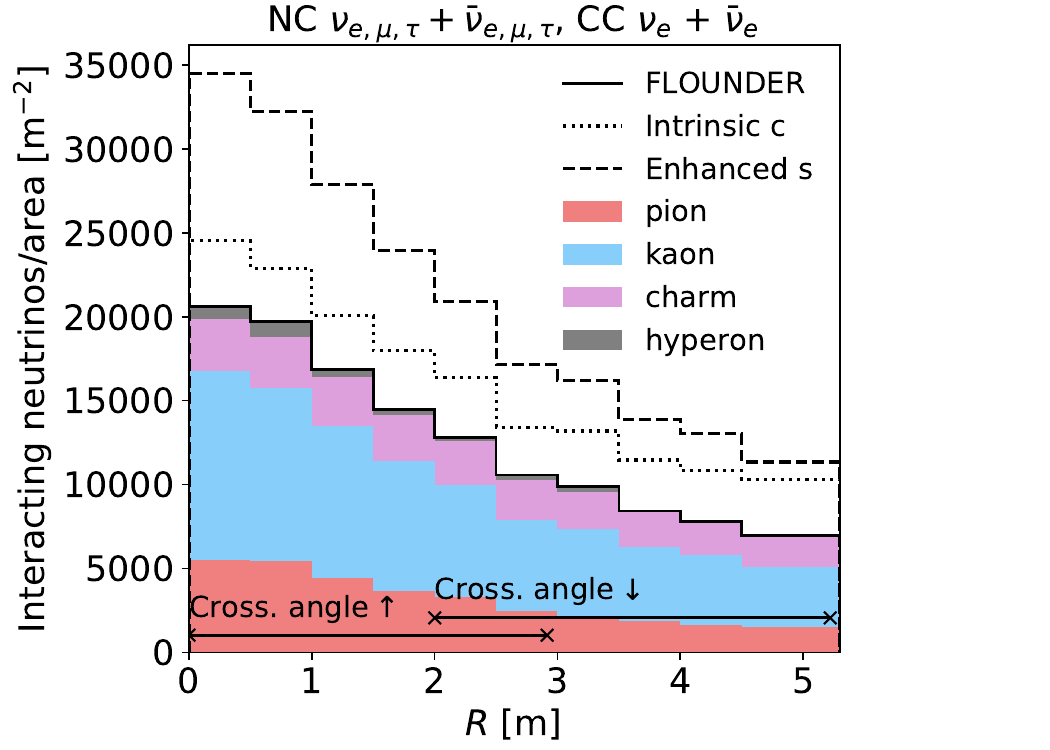}
\caption{The parent hadron composition of the radial spectra of $\nu_e+\overline{\nu}_e$ (left) and $\nu_\mu+\overline{\nu}_\mu$ (middle) CC neutrino interactions as well as cascade-like events (right) in FLOUNDER. The spectra have been normalized by the radial bin transverse area within the detector. Changes of the energy spectrum caused by forward strangeness enhancement and intrinsic charm are shown as a dashed and dotted line, respectively. At 9~km from the IP, the radial distances 
$R\in\{1,2,3,4,5\}$~m 
correspond approximately to the pseudorapidities 
$\eta\in\{9.8,9.1,8.7,8.4,8.2\}$.}
\label{fig:radialSpecta}
\end{figure*}

The sources of systematic uncertainty considered are the resolutions in measuring the muon momentum and angle, as well as the hadronic energy, as summarized in \cref{table:detectorCapabilities}. Moreover, the study of the strange quark PDF at FLOUNDER would rely on identifying a second energetic muon resulting from the decay of a charm quark in the hadronic system. Hence the number of events that can be used to constrain the strange PDF is reduced to approximately $\sim15$\% of all $\nu_\mu$ CC events involving charm production. In conclusion, the detector dimensions and capabilities considered in this work will not suffice for significantly reducing PDF uncertainties without further development and improvement of detector technologies to reduce the systematic uncertainties. We also note that LED will not have sufficient statistics to contribute to PDF measurements. 

\subsection{Neutrino Flux Measurements}
\label{subsec:NeutrinoFluxMeasurements}

\begin{figure}[t]
\centering
\includegraphics[width=0.32\textwidth,trim={0mm 0mm 35mm 0mm},clip]{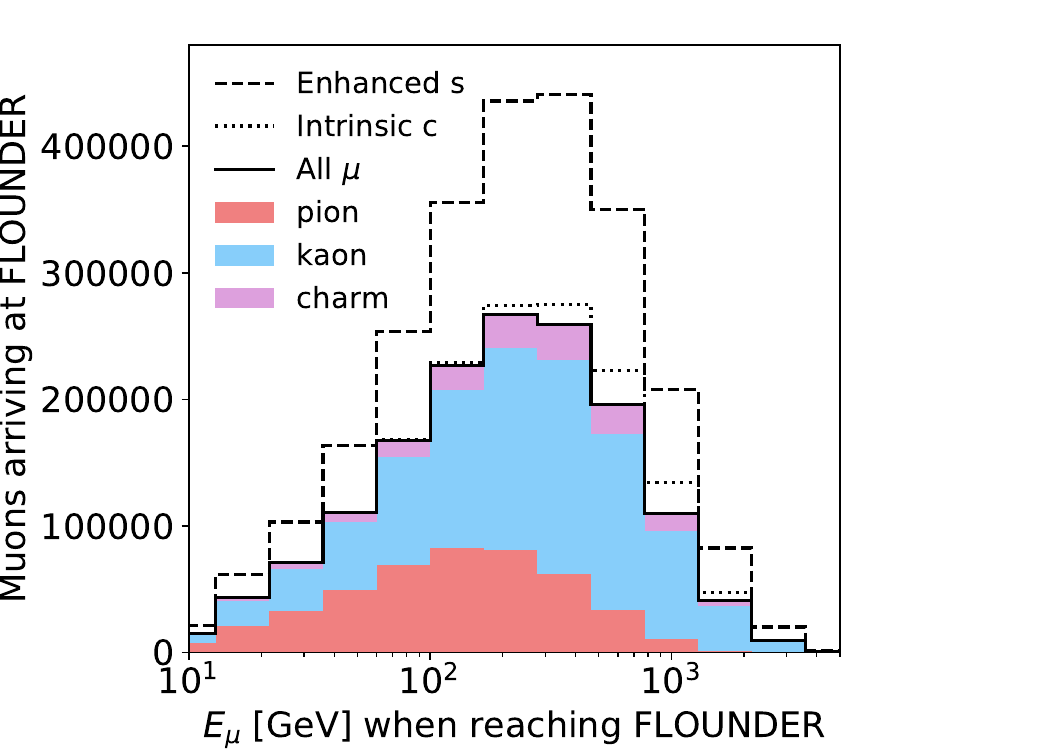}
\caption{Energy spectrum and parent hadron composition of muons from muon neutrino interactions in the rock before FLOUNDER as they enter the detector volume. Changes of the energy spectrum caused by forward strangeness enhancement and intrinsic charm are shown as a dashed and dotted line, respectively.}
\label{fig:muons}
\end{figure}

The LHC's neutrino beam originates from the decay of the lightest hadrons of a given flavor, most importantly charged pions, kaons and charm mesons. Notably, the production of these particles in the forward direction has not been measured before. Therefore, neutrino flux measurements at the LHC provide a novel method to investigate forward particle production and constrain the underlying physics. 

The contributions of various parent hadrons to the neutrino spectra observed at FLOUNDER are shown in \cref{fig:composition_FLOUNDER} as a function of neutrino energy and in \cref{fig:radialSpecta} in terms of radial distance from the beam center. Note that FLOUNDER is expected to obtain considerable event rates with both IP5 crossing angle configurations, effectively increasing the extent of the probed rapidity region beyond the nominal detector dimensions. Generally, neutrinos from light hadron decays tend to have a smaller transverse momenta, and are therefore more collimated around the beam axis, than neutrinos from charm hadron decays, which are more spread out. Pions only contribute to the muon neutrino flux. Kaons contribute mostly to the flux of high energy muon neutrinos, as well as low-energy and low-radial distance electron neutrinos. High-energy and high-radial distance electron neutrinos result predominantly from the decays of charmed hadrons. Given the radial and energy dependence of the different parent hadron contributions to the resulting neutrino flux, the relatively large transverse area of the detector and its sufficiently good spatial and energy resolution make FLOUNDER useful for understanding different parent hadron contributions to the neutrino spectra. 

As discussed in \cref{sec:spectra} and shown in \cref{fig:muons}, muons entering FLOUNDER arise mainly from muon neutrinos produced in light hadron decays which subsequently interact via CC in front of FLOUNDER. Observing these muons thus provides an additional handle to constrain the forward hadron flux. The fraction of neutrinos from kaon decays is increased, in comparison to neutrinos interacting in FLOUNDER, especially at lower energies. This is because neutrinos from kaon decay are on average more energetic and therefore the produced muons can travel further through the rock, effectively increasing the target volume. This also illustrates that the measurement provides complementary information that may help in breaking degeneracies.

The production of forward pions and kaons (and thus the bulk of forward muon neutrinos) is described by hadronic interaction models, subject to sizable modeling uncertainties. This is mainly due to the lack of high-energy forward particle production data, coming only from the neutral pion and neutron measurements by LHCf~\cite{LHCf:2017fnw, LHCf:2018gbv}. Neutrino measurements at the LHC will add complementary data on the forward production of charged pions, charged kaons and neutral kaons. Constraining high energy forward particle production and improving hadronic interaction models is particularly interesting for astroparticle physics, where they are used for simulating particle production in extreme astrophysical systems, as well as cosmic ray interactions in the atmosphere. Notably, for the latter, there is a long-standing discrepancy between the number of muons observed in high-energy cosmic ray air shower observations and the number predicted by hadronic interaction models, which is known as the muon puzzle~\cite{PierreAuger:2014ucz, PierreAuger:2016nfk, EAS-MSU:2019kmv, Soldin:2021wyv, PierreAuger:2024neu}. This problem prevents a measurement of the mass composition of the cosmic ray flux, which is needed for distinguishing different hypotheses on their origins. Extensive studies suggest that this discrepancy is caused by a mismodeling of soft QCD effects in forward particle production at center-of-mass energies above the TeV scale~\cite{Ulrich:2010rg, Albrecht:2021cxw}. Further studies have shown that this problem could be resolved through an enhanced rate of strangeness production in the forward direction~\cite{Allen:2013hfa, Anchordoqui:2016oxy, Anchordoqui:2019laz}. Since kaon decays are one of the main sources of neutrinos at the LHC, measurements of collider neutrino fluxes will allow to constrain and test these scenarios, ultimately helping to resolve the muon puzzle. A phenomenological model for enhanced strangeness production has been introduced in Ref.~\cite{Anchordoqui:2022fpn, Sciutto:2023zuz}. As illustrated by the dashed line in \cref{fig:composition_FLOUNDER,fig:radialSpecta,fig:muons}, this model predicts a significant increase of the neutrino event rate that can be tested with FLOUNDER. 
It should be noted that although FASER$\nu$ Run~3 data will already constrain the enhanced strangeness scenario preferred in Ref.~\cite{Anchordoqui:2022fpn}, it is possible that the effect is less pronounced in $pp$ collisions at the LHC~\cite{Anchordoqui:2016oxy,Anchordoqui:2022fpn} or exhibit non-trivial energy and rapidity dependence~\cite{Sciutto:2023zuz}. Accounting for such cases will require the additional event rates offered by FLOUNDER or the FPF experiments~\cite{Kling:2023tgr}.

In contrast to light hadrons, forward charm production can, in principle, be described by perturbative QCD. Charm quarks are dominantly produced via gluon fusion, where one gluon carries a large momentum fraction $x\sim 1$ while the other carries a very small momentum fraction $x \sim 4 m_c^2 / s \sim 10^{-7}$. The neutrino flux from charm decays is therefore sensitive to both high-$x$ physics, such as intrinsic charm~\cite{Maciula:2022lzk}, and low-$x$ physics, especially to the gluon PDF in an uncharted kinematic regime around $x\sim 10^{-7}$. Such measurements will allow studying novel QCD phenomena, including BFKL dynamics and the onset of gluon recombination~\cite{Bhattacharya:2023zei}. Forward charm measurements will also provide useful input for astroparticle physics, for instance for estimating the prompt atmospheric neutrino flux~\cite{Jeong:2023gla}. As an example, the effect of an intrinsic charm component on the resulting neutrino flux, following the BHPS model~\cite{Brodsky:1980pb} implemented in the CT14 PDF~\cite{Hou:2017khm} as estimated in Ref.~\cite{Maciula:2022lzk}, is shown as a dotted line in \cref{fig:composition_FLOUNDER,fig:radialSpecta,fig:muons}. With this model, there is a significant increase in the number of electron neutrinos at high energies, as well as a modest increase in the number of muon neutrinos at high energies.

\begin{figure*}[t!]
\centering
\includegraphics[width=0.32\textwidth,trim={0mm 0mm 35mm 0mm},clip]{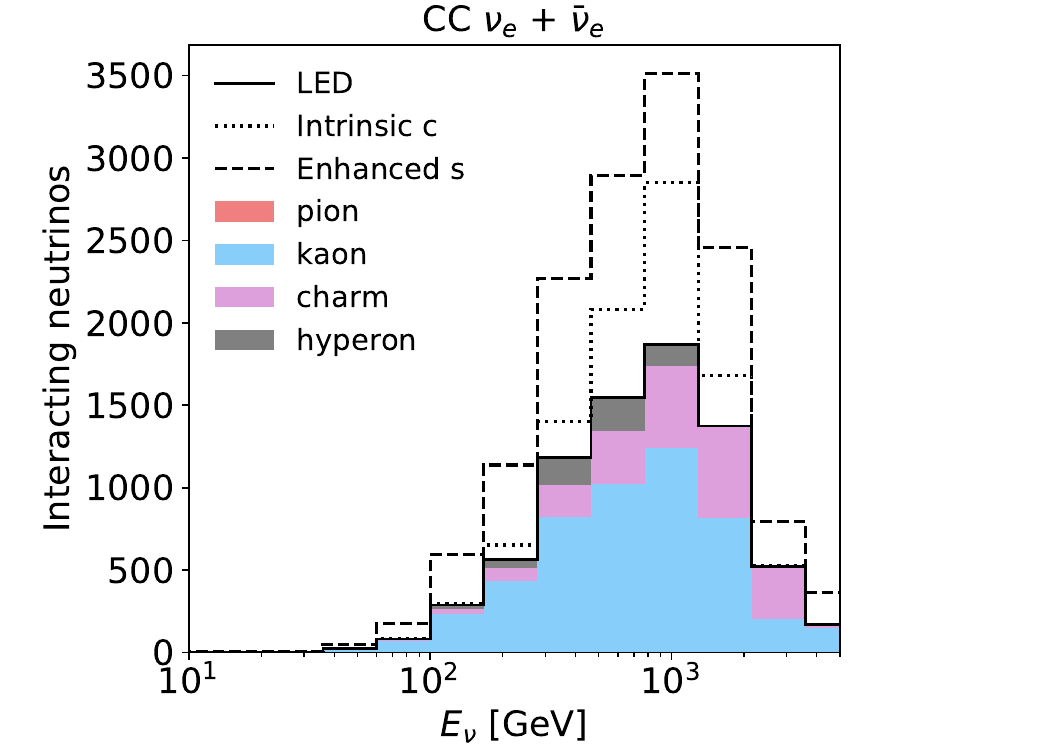}
\includegraphics[width=0.32\textwidth,trim={0mm 0mm 35mm 0mm},clip]{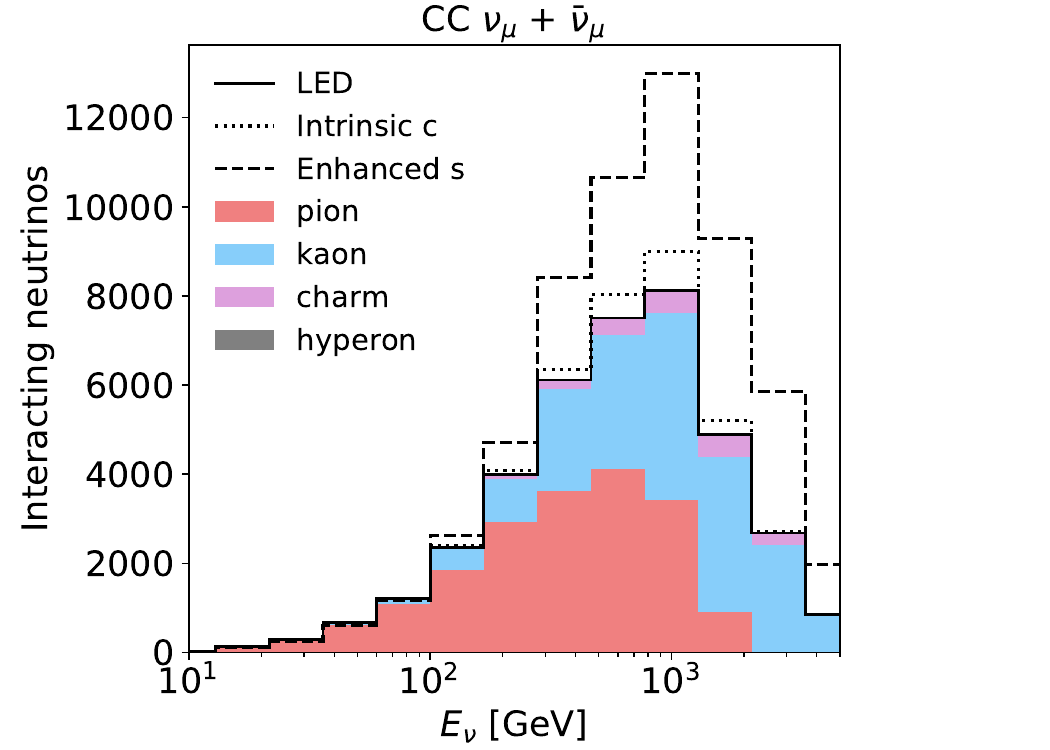}
\includegraphics[width=0.32\textwidth,trim={0mm 0mm 35mm 0mm},clip]{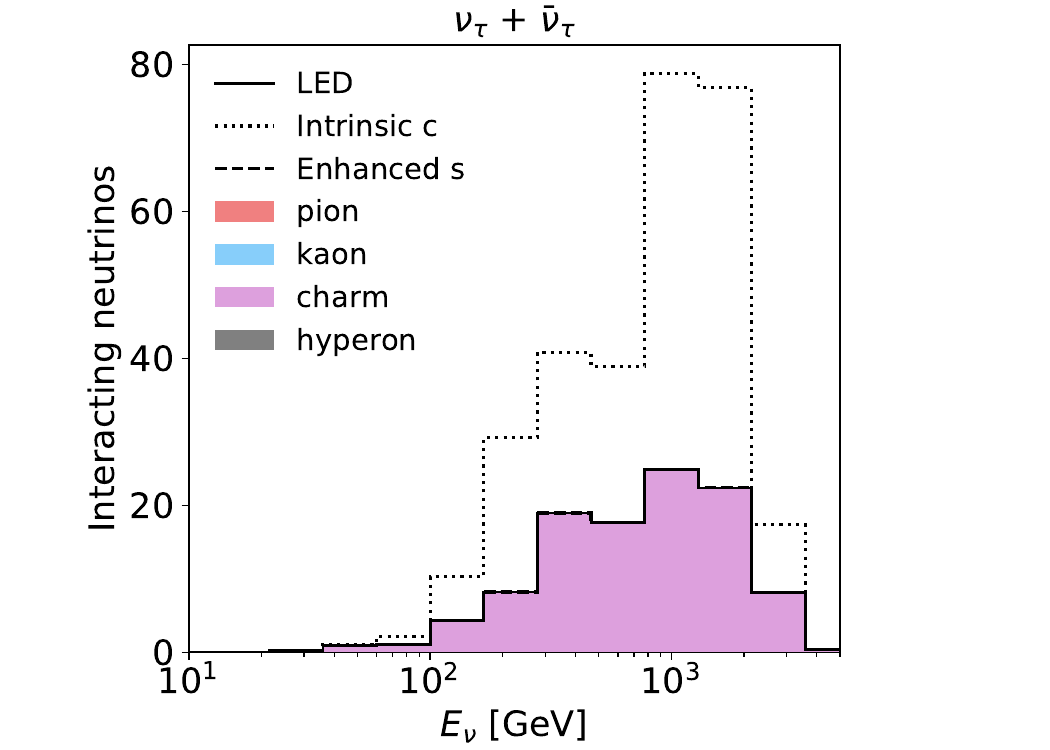}
\caption{The parent hadron composition of the spectra of $\nu_e+\overline{\nu}_e$ (left), $\nu_\mu+\overline{\nu}_\mu$ (center) and $\nu_\tau+\overline{\nu}_\tau$ (right) CC neutrino interactions in a 200 ton LED in terms of the incoming neutrino energy. Changes of the energy spectrum caused by forward strangeness enhancement and intrinsic charm are shown as dashed and dotted line, respectively. See text for details.   
}
\label{fig:composition_LED}
\end{figure*}

Accessing this physics potential requires measuring the energy and angular spectra of neutrinos for different flavors. It may, however, be difficult to reliably distinguish a high energy electromagnetic shower from a hadronic shower in a water Cherenkov detector. Therefore, the experimental signature of a $\nu_e$ CC event at FLOUNDER can be closely mimicked by NC events induced by any neutrino flavor. Although methods for reliably distinguishing $\nu_e$ CC events in a water detector have been developed and discussed at lower energy ranges~\cite{CHIPS:2024vpb, Tingey:2022evd} and their observation at higher energies could be possible due to the Landau-Pomeranchuk-Migdal effect~\cite{Landau:1953ivy, Landau:1953gr, Migdal:1956tc}, it will require further detector simulation work to determine the accuracy up to which this can be done at TeV energies. It is therefore instructive to consider both an optimistic case, in which the $\nu_e$ CC rates can be measured directly, as well as a conservative scenario, in which only cascade-like interactions arising from both $\nu_e$ CC and $\nu_\ell$ NC (with $\ell\in\{e,\mu,\tau\})$ can be identified. These two scenarios are shown for FLOUNDER in \cref{fig:composition_FLOUNDER,fig:radialSpecta}. Comparing the two possible cases for FLOUNDER, we can see that the information on electron neutrinos is somewhat diluted in the pessimistic scenario due to the presence of muon neutrino NC events. The spectrum of cascade-like events contains a sizable component of $\nu_\mu$ NC, making it more similar in its hadronic composition to the $\nu_\mu$ CC spectrum which has a subdominant charm contribution in contrast to the $\nu_e$ flux. This illustrates that, in order to obtain the best understanding of the neutrino flux composition and gain better access to forward charm production, a detector should ideally be able to distinguish $\nu_e$ CC from NC events. \medskip

The flux composition of CC neutrino interactions that could be observed at a 200~ton lake emulsion detector is shown in \cref{fig:composition_LED}. Unlike for FLOUNDER, an emulsion detector can reliably identify not only electron and muon CC neutrino interactions, but also tau neutrinos interactions, which are only produced in charm hadron decays. Despite the lower event rate, in comparison to FLOUNDER, this offers an additional handle to constrain the underlying models of forward particle production, in particular effects associated to forward charm production. 

\subsection{Dark Sectors}

As pointed out in Ref.~\cite{Feng:2017uoz}, the forward region of the LHC may also produce large numbers of new particles that are light and very weakly interacting. Such particles have been proposed to address many of the outstanding questions in particle physics, for example to explain the nature and observed abundance of dark matter, neutrino masses and the observed matter-antimatter asymmetry in the universe. Several experiments have been proposed to exploit this opportunity. This includes FASER, which searches for the decay of long-lived particles~\cite{FASER:2018ceo, FASER:2018eoc, FASER:2018bac, FASER:2019aik, FASER:2022hcn}; FLArE, proposed to look for scattering light dark matter~\cite{Batell:2021blf, Batell:2021aja, Batell:2021snh}; and FORMOSA, proposed to search for millicharged particles~\cite{Foroughi-Abari:2020qar}. 

Both light dark matter scattering and millicharged particles lead to only small, typically sub-GeV, energy deposits in the detector. While those are in principle detectable at a water Cherenkov detector, these signatures may suffer from sizable low-energy backgrounds induced by the neutrino beam or cosmic rays. A careful study of these backgrounds and the detector's capability to detect the signal, which require a full detector simulation, would be needed, but is beyond the scope of this work.

The situation is different for long-lived particles, which can decay inside the detector volume and deposit TeV energies. However, while providing a spectacular signal, neutrino interactions occurring in the target volume pose a potential source of background.

\begin{figure*}[t!]
\centering
\includegraphics[width=0.46\textwidth]{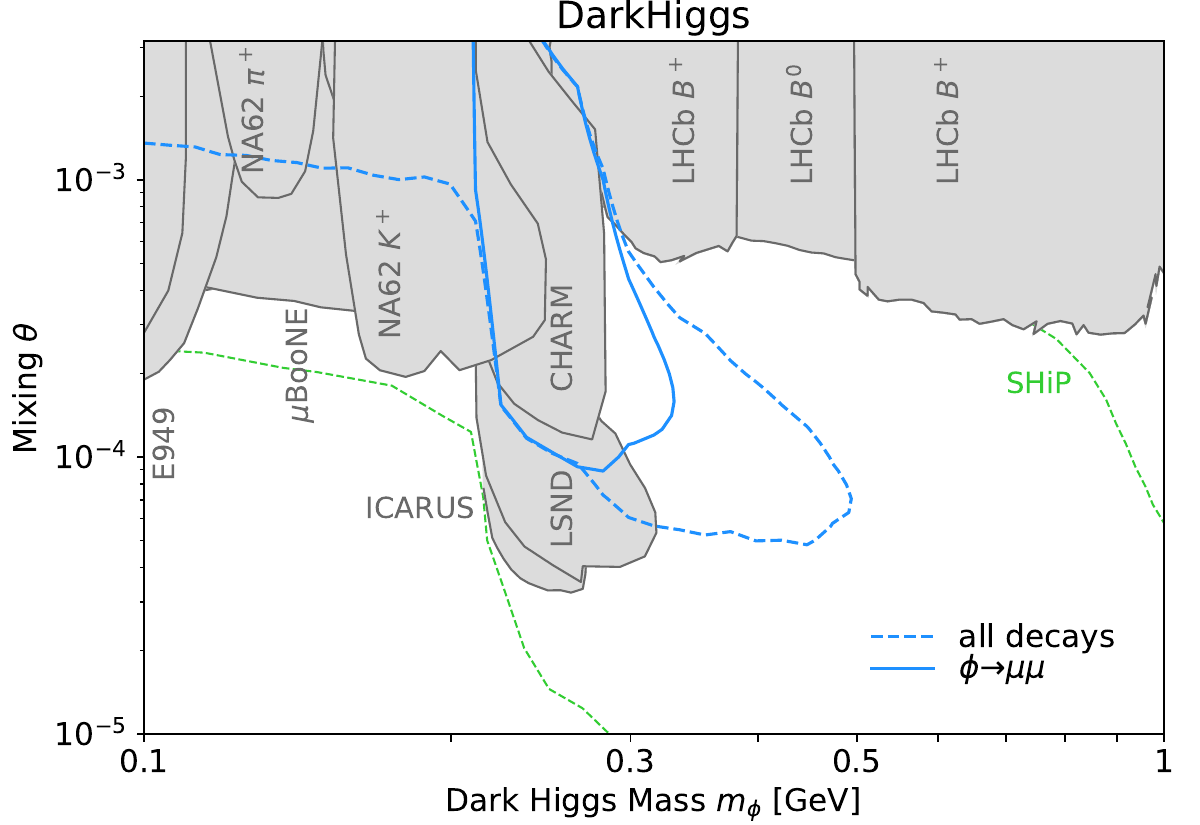}
\includegraphics[width=0.46\textwidth]{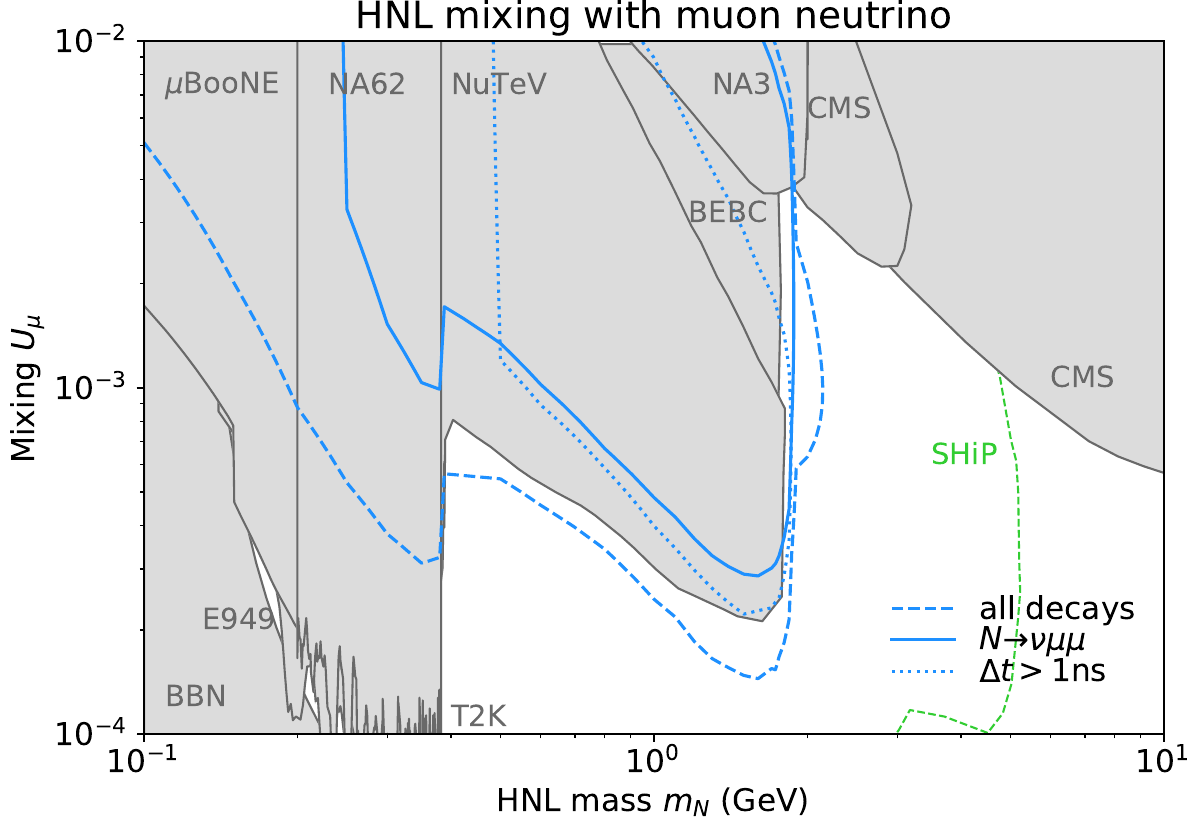}
\caption{Landscape of the Dark Higgs (left) and HNLs (right). Existing constraints are shown in gray. The blue lines enclose the region in which more than three signal events are expected, taking into account all decays (dashed), only decays into muons (solid) and decays that are delayed by more than 1~ns (dotted, right figure only). These contours do not account for signal efficiencies and potential background, which can be sizable and are expected to reduce the sensitivity. 
The SHiP bounds are obtained from Ref.~\cite{Fernandez:2025dcy}.
}
\label{fig:LLPs}
\end{figure*}

Light LLPs primarily decay into pairs of electrons or photons. These then interact and initiate an energetic electromagnetic shower in the detector. This signature has to be distinguished from CC electron neutrino interactions, which could, in principle, be done using the presence of an additional recoil hadronic shower. Neutrino interactions, however, have a roughly flat inelasticity distribution, $d\sigma/dy \approx \text{const.}$, where $y$ is the fraction of the energy transferred from the neutrino to the hadronic system. For example, in roughly 1\% of all $\nu_e$ CC interactions the electron gains more than 99\% of the energy while the hadronic system is very soft and carries less than 1\% of the energy. In this case, only the electromagnetic shower would be visible. Since we expect about a hundred thousand $\nu_e$ CC interactions inside FLOUNDER, this corresponds to about 1000 background events. While dedicated simulations will provide more accurate background estimates, this argument suggest that LLPs decays into electrons or photons will likely suffer substantial backgrounds. 

Heavier LLPs may also decay into pairs of muons, which would be visible as long tracks in the detector. This signature requires the ability to identify and separate the two muons, which would be challenging given that the muon tracks are highly collimated. Backgrounds could arise from charm associated CC muon neutrino interactions, in which the charm hadron decays to a muon, which could be suppressed by vetoing the presence of an (even relatively soft) hadronic shower. A likely irreducible background arises from neutrino trident production $\nu_\mu N \to \nu_\mu \mu \mu N$. Following Ref.~\cite{Altmannshofer:2024hqd}, a few such events are expected to occur inside the detector\footnote{The statistics at FLOUNDER are however expected to be insufficient for conclusive trident observations, as the main task of contemporary trident studies is to assess the backgrounds producing similar experimental signatures, e.g. diffractive charm production and particles imitating muon tracks, and to place cuts on them without diminishing the signal event rates.}. Nevertheless, LLP decays into muons present the most promising channel. 

One may also consider LLP decays into muons in the rock before FLOUNDER. The two muons could then enter the decay volume, providing an alterative way to detect the signal. However, this signal could be mimicked by charm associated muon neutrino CC interactions, in which the produced charm hadron decays into a second muon. We have estimated the associated rate of such events using the simulation described at the end of \cref{sec:spectra} and find that charm associated muon neutrino interactions in the rock will lead to a sizable rate of muon pairs entering FLOUNDER, and requiring both muons to have energies above 100~GeV (500~GeV) reduces this rate to about 600 (20) events. These cuts would, however, also significantly reduce the signal efficiency. We conclude that this signature will not allow for a background free search. 

LLPs may also decay into two or more hadrons, which would initiate a hadronic shower. This is very similar to a neutral current neutrino interaction, with no obvious handle to separate them. Since we expect about a hundred thousand such neutrino interactions, detecting hadronically decaying LLPs seems impossible.

From the above considerations, it is clear that FLOUNDER will be primarily sensitive to LLPs decaying to muons. To investigate FLOUNDER's potential to probe dark sectors, we consider two benchmark models permitting this decay: the dark Higgs boson and a heavy neutral lepton (HNL) mixing with the muon neutrino. For both cases, we use the FORESEE simulation package~\cite{Kling:2021fwx} to estimate the expected number of LLP decays inside the detector volume. When estimating the flux, we use the dedicated forward physics tune of \texttt{Pythia}~\cite{Fieg:2023kld} to simulate light hadron production and the particle spectra obtained in Ref.~\cite{Buonocore:2023kna} using \texttt{POWHEG} and \texttt{Pythia} for heavy hadron production. 

The dark Higgs $\phi$ is a new scalar that mixes with the SM Higgs field, thereby obtaining Higgs-like couplings to all SM particles~\cite{Feng:2017vli}. Its low-energy Lagrangian is given by $ \mathcal{L} = -m_\phi^2 \phi^2 - \sin \theta \sum_f y_f \phi \bar f f $, where $m_\phi$ denotes its mass and $\theta$ the mixing. For the modeling of production and decay, we follow the phenomenological description of Ref.~\cite{Winkler:2018qyg}. The dark Higgs is mainly produced in B-hadron decays and mostly decays to the heaviest kinematically accessible final states: electron pairs for $m_\phi<2m_\mu$, muon pairs for $2m_\mu < m_\phi < 300~\mev$ and hadrons for higher masses. 

Results for the dark Higgs are shown in the left panel of \cref{fig:LLPs}. The gray shaded regions represent existing constraints from searches at LHCb~\cite{LHCb:2015nkv, LHCb:2016awg}, CHARM~\cite{Winkler:2018qyg}, LSND~\cite{Foroughi-Abari:2020gju}, NA62~\cite{NA62:2020xlg, NA62:2020pwi}, E949~\cite{BNL-E949:2009dza}, MicroBoone~\cite{MicroBooNE:2021sov} and ICARUS~\cite{ICARUS:2024oqb}, while the green dotted line indicates the projected bound for the recently-approved SHiP experiment~\cite{Fernandez:2025dcy}. The blue dashed line indicates the region of parameter space within which more than three dark Higgs decays are expected to occur inside the FLOUNDER volume. Under the assumption of perfect signal efficiency and negligible backgrounds, this would correspond to the discovery reach. It therefore indicates the ultimate sensitivity that could, in principle, be obtained with a detector at this location. We can see that it extends beyond existing constraints for masses between 300 and 500~MeV. The blue solid line shows the three event contour after requiring the dark Higgs to decay into muon pairs. This reduces the event rate at higher masses, but potential sensitivity to unprobed regions of parameter space remain. We note, however, that this contour does not account for potential background or detection inefficiencies that will likely further reduce the sensitivity. 

The HNL $N$ is a new neutral fermion that mixes with the active neutrinos, in this case the muon neutrino. Its phenomenology is described by its mass $m_N$ and mixing $U_\mu$. At the forward direction of the LHC, HNLs would primarily be produced via weak meson decays, most importantly kaons, charm and beauty hadrons, and can decay through either NC or CC interaction~\cite{Kling:2018wct}. We use the HNLCalc package~\cite{Feng:2024zfe} to describe all relevant production and decay modes. 

The right panel of \cref{fig:LLPs} shows the HNL, with the gray regions representing existing constraints from BEBC~\cite{WA66:1985mfx}, CMS~\cite{CMS:2022fut, CMS:2024ake}, E949~\cite{E949:2014gsn}, MicroBoone~\cite{MicroBooNE:2023eef}, NA3~\cite{NA3:1986ahv}, NuTeV~\cite{NuTeV:1999kej}, NA62~\cite{NA62:2021bji}, T2K~\cite{T2K:2019jwa}, and BBN~\cite{Sabti:2020yrt}, as provided by the Heavy Neutrino Limits package~\cite{Fernandez-Martinez:2023phj}. The green dotted line represents the SHiP bound~\cite{Fernandez:2025dcy}. As before, the blue dashed line corresponds to more than three expected decay events inside the detector volume, indicating the theoretical upper limit on the sensitivity of such a detector: new regions of parameter space are probed for masses between 400~MeV and 2.5~GeV. Restricting the search to muons pairs almost entirely diminishes this sensitivity even before considering backgrounds or inefficiencies. Another handle to suppress backgrounds is provided by timing: due to the long distance between the LHC collision point and the detector, heavy HNLs can arrive with a substantial delay. This would allow to distinguish their decays from neutrino interactions, which travel with the speed of light. The blue dotted line indicates the potential of such a search by requiring a delay $\Delta t > 1$~ns. \medskip
 
Overall, these examples have illustrated that a reasonable sized detector at the exit points of the LOS has, in principle, potential to probe dark sectors through LLP decays. However, in the examples mentioned above, the sensitivity does not greatly exceed existing constraints, even before considering backgrounds or inefficiencies, and is significantly weaker than that of the recently approved SHiP experiment~\cite{Fernandez:2025dcy}. In addition, other experiments, including FASER operating during the HL-LHC era, are also sensitive to similar regions of parameter space~\cite{FASER:2018eoc}.

\section{Conclusions}
\label{sec:conclusions}

While the LHC was primarily built to search for the Higgs boson and other particles at and beyond the  electroweak scale, it is also the source of the most energetic neutrinos produced by human kind. These neutrinos have recently been observed for the first time by the FASER and SND@LHC experiments, thereby initiating a new field of collider neutrino physics. Both detectors sit in underground locations, close to the ATLAS IP, and larger experiments in a purpose-built cavern have been proposed in the context of the FPF. In this publication we consider the possibility of not placing neutrino detectors underground, but at the surface exit points of the neutrino beam. 

The topographic study presented within this work has located the emergence points of the neutrino beams originating at the LHC. The furthest of them is at a distance of more than 100~km, and most of them over 20~km, away from the IP. The sites east of IP5, 9~km away at the bottom of Lake Geneva, and west of IP5, 19~km removed in the Jura mountains, are identified as the most promising candidates for surface level detectors designed for a neutrino and dark sector program. 
At these locations, the spread of the neutrino flux over large distances requires considering kiloton detectors, necessitating reliance on technologies with smaller cost-per-volume than those utilized in the comparably small detectors close to the IP.
However, this limits tracking granularity and particle identification abilities, which are required for an important part of the physics program of the proposed and existing underground detectors.

We have considered several possible detector technologies at both locations. The site at the bottom of Lake Geneva is suitable for e.g. a 20 to 200~ton submerged tungsten-emulsion detector or a 1.8~kiloton water Cherenkov detector. During the high-luminosity LHC run, the 20 (200) ton lake emulsion detector would yield approximately 40\% (4 times) the number of neutrino interaction events expected to be collected by the 1.1~ton FASER$\nu$ 480~m away from IP1 already during Run~3 of the LHC. In comparison to detectors closer to the IP, the lake location nonetheless offers a reduced forward muon background, and an emulsion detector in the lake would allow for a long exposure time for a large quantity of emulsion, providing about a hundred tau neutrino events for 200~ton detector masses. 
Of the surface-level detectors, the highest total event rates for the HL-LHC run are however predicted at the 1.8~kiloton water Cherenkov detector.
They are more than 40 times the FASER$\nu$ Run 3 expectation, although remain at less than 40\% of the HL-LHC run levels of the proposed 20~ton FASER$\nu$2 detector to be hosted at the FPF. 
Obtaining similar numbers at IP5W requires a 15~kiloton detector, which would be similar in size to the NOvA far detector. 
This implies that obtaining event rates comparable to those expected at the underground near detector locations requires considerably larger detectors at the exit point distances, and that the best opportunity for a surface-level experiment would be provided by the emergence point in Lake Geneva.

The determination of the lepton identification capabilities and systematic uncertainties of a water Cherenkov detector requires further detector simulations outside the scope of this work, which therefore opts for conservative uncertainty estimates based on existing and proposed water detector designs. There will be no charge identification, and it is unclear if $\nu_e$ CC events can be separated from NC. The present work mostly relies on processes with final state muons, and assumes rough energy and angular resolution.

The statistics expected at the water Cherenkov detector suffice for constraining several physics scenarios within and beyond the SM, such as forward hadron production and cross section measurements, although without $\nu$/$\overline{\nu}$ identification, as well as testing the enhanced strangeness scenario proposed to solve the cosmic ray muon puzzle. However, with the assumed lepton identification capabilities and uncertainty estimates, the detector provides limited insight into nucleon structure, and would only provide a small increase in sensitivity for select models of light long-lived particles. 
Full consideration of all decay modes nonetheless necessitates a detector design for which electron channels can be separated from similar experimental signatures, requiring further study.

The timeline for implementing any of the conceptual detectors discussed depends on details of their technical design not available yet. It should be noted that the current schedule has the HL-LHC starting in summer-2030, with significant luminosity production not starting until a year later, giving more than five years for the detectors to be implemented before the start of HL-LHC operations. Since the proposed detectors are far from the LHC, they can be implemented independently of the LHC schedule. If they start physics operation after the start of the HL-LHC the event rates discussed in this paper will, of course, scale with the collected luminosity. For the physics measurements discussed, collecting half of the luminosity used in the projections (3/ab) will not have a large impact on the physics message.

The prospect of taking advantage of the surface-level exit points of the LHC neutrino beams is a novel research direction distinct from the existing investigations of very energetic astrophysical neutrinos or low-energy accelerator neutrinos. 
Although our studies indicate that the FPF and FASER physics programme cannot be replaced by surface-level detectors at the LHC, this may motivate further simulation work on TeV neutrino interactions in water Cherenkov detectors, in order to refine its physics potential. Finally, to study the feasibility of this further, a modest programme to reduce the uncertainty of the LOS exit point locations for the preferred exit points would be valuable. This should include reducing the uncertainties associated with the depth of the lakebed as well as characterizing possible time variation in the depth.

\section*{Acknowledgements}

We thank Jonathan Feng, Roshan Mammen Abraham and Kohei Chinone for useful discussions. We also thank Gianluigi Arduini, Angelo Infantino and the CERN Physics Beyond Colliders study group. TM thanks the fellow authors of Ref.~\cite{Cruz-Martinez:2023sdv}, acknowledging the use of the PDF profiling pipeline developed for the study. We note that Nicholas Kamp and collaborators have also considered the option to place a neutrino detector in Lake Geneva, with preliminary results presented at the 31st International Conference on Neutrino Physics and Astrophysics~\cite{Kamp2024Neutrinos} and at a CERN joint BSM/FPC meeting~\cite{Kamp:2024pbc}. We are also grateful to the authors and maintainers of many open-source software packages, including
\texttt{CRMC}~\cite{crmc201},
\texttt{RIVET}~\cite{Buckley:2010ar, Bierlich:2019rhm},
\texttt{scikit-hep}~\cite{Rodrigues:2019nct}.

The work of F.K.~was supported by the Deutsche Forschungsgemeinschaft under Germany's Excellence Strategy -- EXC 2121 Quantum Universe -- 390833306. 
T.M. is supported in part by U.S.~National Science Foundation Grants PHY-2111427 and PHY-2210283 and Heising-Simons Foundation Grant 2020-1840. 
The work of M.F.~was supported in part by U.S. National Science Foundation Grants PHY-2111427 and PHY-2210283, as well as by NSF Graduate Research Fellowship Award No. DGE-1839285.
S.W.B supported by National Science Foundation Grant PHY-2411780. 
A.A. is supported by the European Research Council (ERC) under the European Union’s Horizon 2020 research and innovation programme (Grant agreement No. 101002690) and JSPS KAKENHI Grant No. 23H00103.

\bibliography{references}

\end{document}